\documentclass[twocolumn,epjc3]{svjour3}          

\usepackage{graphicx}
\usepackage{dcolumn}
\usepackage{bm}
\usepackage{amsmath,multirow}
\usepackage{amssymb}
\usepackage{subfigure}
\usepackage{color}
\usepackage{verbatim} 
\usepackage{float}
\graphicspath{{./}{figpdf/}}

\RequirePackage[T1]{fontenc}

\smartqed  

\journalname{EPJC}

\begin{document}

\title{Chaos of charged particles near a renormalized group
improved Kerr black hole in an external magnetic field}


\author{
Junjie Lu  
\and Xin Wu$^{a}$} 

\thankstext{e2}{e-mail: wuxin$\_$1134@sina.com; 21200006@sues.edu.cn (corresponding author)}

\institute{School of Mathematics, Physics and Statistics, Shanghai
University of Engineering Science, Shanghai 201620, China
}

\date{Received: date / Accepted: date}

\maketitle

\begin{abstract}

In a quantum theory of gravity, a renormalization group improved
Kerr metric is obtained from the Kerr metric, where the Newton
gravitational constant is modified as  a function of the radial
distance. The motion of neutral test particles in this metric is
integrable. However, the dynamics of charged test particles is
nonintegrable when  an external asymptotically homogeneous
magnetic field exists in the vicinity of the black hole. The
transition from regular dynamics to chaotic dynamics is
numerically traced as one or two dynamical parameters vary. From a
statistical point of view, the strength of chaos is typically
enhanced as both the particle energy and the magnetic field
increase, but it is weakened with increasing the particle angular
momentum and the black hole spin. In particular, an increase of
the quantum corrected parameter weakens the extent of chaos. This
is because the running Newton gravity constant effectively weakens
the central gravitational attraction and results in  decreasing
sensitivity to initial conditions.


\end{abstract}

\section{Introduction}

Curvature singularities inevitably exist in several classical
general relativistic  black hole solutions, including
Schwarzschild black holes, Reissner-Nordstr\"{o}m black holes,
Kerr black holes and Kerr-Newman black holes. In the near regions
of such singularities, the classical theory of general relativity
becomes useless. In this case, some other theories of gravity are
necessary to avoid the singularities in the black hole solutions.
The so-called regular black hole solutions satisfy this
requirement. Bardeen black holes  [1] are the first regular  black
hole solutions but are not exact solutions to Einstein equations
in general relativity. The first regular exact black hole solution
of the Einstein field equations coupled to a nonlinear
electrodynamics was derived by Ay\'{o}n-Beato and Garc\'{i}a [2].

Quantum theories of gravity are also possible to provide
non-singular black hole solutions. Based on the methods of the
standard quantum field theory and the techniques of the functional
renormalization group, renormalization group improved (RGI)
Schwarzschild or Kerr black hole solutions were given in
literature, e.g. [3-5]. In the framework of the asymptotic safety
quantum gravity, these RGI black hole solutions simply arise from
the classical black hole solutions by modifying the Newton
gravitational constant as varying position-dependent functions. In
this way, the RGI approach resolves the issue of black hole
singularities, i.e., completely avoids the formation of
singularity during the gravitational collapse of a dying star.
More significantly, it leads to the existence of a minimal black
hole mass and the formation of a cold Planck-scale remnant. It
offers a possible resolution to the information loss paradox by
preventing complete Hawking evaporation and bridges the gap
between semiclassical gravity and full quantum gravity. In the
standard semiclassical scenario, the Hawking radiation decreases
the mass of a black hole but increases the Hawking temperature.
Doesn't this process stop before the entire mass of the black hole
is completely emitted? As far as the problem on the final stage of
Hawking evaporation is concerned, the RGI approach argues that the
Hawking evaporation continues until the mass of the black hole
approaches a certain critical mass with the order of the Planck
mass [3].

The quantum effects of gravity play a crucial role in not only
addressing  the limitation problems of Einstein's general
relativity but also exerting influence on the dynamics of test
particles or photons near quantum-improved black holes and
physical properties of the black holes. In terms of the weak and
strong deflection gravitational lensing of the RGI Schwarzschild
black hole combined with the shadow of M87*, the two quantum
improved  parameters are constrained in Reference [6]. In this
case, it is difficult to distinguish the RGI black hole from the
Schwarzschild black hole. However, scale-dependent Planck stars
can be distinguished from the RGI Schwarzschild black holes with
the aid of the motion of particles around these black holes [7].
It was shown in [8] that the innermost stable circular orbit
radius of a charged particle near the RGI Schwarzschild black hole
surrounded by an external asymptotically uniform magnetic field
decreases as a quantum parameter increases. By use of the
effective potential, circular motions of charged particles around
Schwarzschild or Kerr black holes in external magnetic fields were
investigated in [9,10]. The epicyclic motions of charged particles
near the RGI Schwarzschild black hole immersed in the external
magnetic field are useful to theoretical interpretations of 3:2
twin-peak quasi-periodic oscillation (QPO) in the X-ray power
density spectra of Galactic microquasars [11]. Based on QPO
frequency data from micro-quasars, Kalb-Ramond gravity was tested
in [12]. Although the external magnetic field in the vicinity of
the black hole is too weak to affect the integrability of the RGI
Schwarzschild spacetime  (i.e. the integrability of the motion of
neutral particles near the black hole), it may exert an important
influence on the motion of charged particles near the black hole
and even causes the charged particle motion to be nonintegrable.
In fact, chaos of charged particles around the RGI Schwarzschild
black hole with the external magnetic field  was confirmed in Ref.
[13]. In the constraint of the two  quantum improved  parameters
based on  the shadow of M87* black hole, a small variation of one
parameter brings an almost negligible contribution to a transition
from regular dynamics to chaotic dynamics, but a small increase of
another parameter weakens the extent of chaos. The existence of
chaotic dynamics of charged particles around other black holes
with external magnetic fields was found in many papers, e.g.
[14-18].

 Applying the Newman-Janis algorithm to the RGI
Schwarzschild metric from the exact evolution equation for the
effective average action [3], Torres [4] gave the RGI rotating
black hole metric. The shadows of the RGI rotating black hole
become small and are more distorted when the two quantum
parameters get from small to large, compared with those of the
Kerr black hole [19]. In the infra-red limit, the
running Newton constant has only one parameter under the
asymptotically safe gravity and the RGI rotating black hole metric
becomes the RGI Kerr metric [5]. It was also shown that the
ergosphere increases with the parameter increasing. In addition,
the efficiency of Penrose process is larger in the RGI Kerr black
hole than in the Kerr black hole when the two black holes have the
same spins. The author of [20] considered the quantum gravity
modification to the effect of a thin accretion disk spiraling near
the RGI Kerr black hole in the infra-red limit of asymptotic safe
gravity theory. It was found that an increase of the quantum
improved  parameter gives rise to decreasing the radius of the
innermost stable circular orbit, lifting the peaks of the
radiation properties of the disk and increasing the accretion mass
efficiency. The obtained results regarding the effects of the
quantum gravity modification on the accretion disk near the RGI
Kerr black hole are almost the same as those near the
RGI-Schwarzschild black hole [21].

The RGI Kerr spacetime [5,20] is more complicated
than the RGI  Schwarzschild one [3] because the rotation of the
black hole causes dragging effects of the spacetime. As was
mentioned above, chaotic dynamics of charged particles around the
magnetized  RGI Schwarzschild black hole and one of the two
quantum improved parameters weakening the chaotic properties  were
shown in [13]. Now, we want to focus on the dynamics of charged
particles near the RGI Kerr black hole [5,20] immersed in an
external magnetic field in the present paper. In particular, we
are interested in surveying the effect of a small change of the
quantum improved parameter on a transition from regular dynamics
of charged particles to chaotic dynamics of  charged particles. In
this way, we desire to know whether the effect of the quantum
improved parameter on chaos under dragging effects of the
spacetime is similar to that of one or two parameters on chaos in
the RGI Schwarzschild case [13]. In addition, an explicit
symplectic integrator, which conserves the symplectic geometric
structure and shows no secular drift in errors of motion constants
during a long-term integration of a Hamiltonian system, was well
employed in [13]. The algorithmic construction is based on a
time-transformed Hamiltonian split into six explicitly integrable
parts. To save computational cost, we will split a
time-transformed Hamiltonian associated to the magnetized RGI Kerr
black hole into three explicitly solvable pieces. 

For the sake of our purposes, we introduce a Hamiltonian system
for the description of charged particle motion near the RGI Kerr
black hole surrounded by the external magnetic field in Sect. 2.
Numerical investigations are given to the dynamics of  charged
particles in Sect. 3. Finally, the main results are summarized in
Sect. 4.

\section{Hamiltonian of charged particle motion near the RGI Kerr black
hole}

In the theory of quantum gravity, Haroon et al. [5] modified the
Newton gravitational constant $G_0$ as  a function of the radial
distance $r$:
\begin{eqnarray}
G(r)&=&G_0(1-\frac{\xi}{r^2}),
\end{eqnarray}
where $\xi$  is a free parameter regarding the contribution of the
quantum effect to the geometry of spacetime. That is, the running
gravitational coupling constant of [3,4] in the infra-red limit of
asymptotic safe gravity theory is the expression of Eq. (1).
Replacing  $G_0$ with $G(r)$ of Eq. (1) and taking  $G_0$ in Eq.
(1) and the speed of light $c$ as geometric units $c=G_0=1$,
Haroon et al. [5] obtained a line element of the RGI Kerr black
hole with mass $M$ in the Boyer-Lindquist coordinates
$(t,r,\theta,\phi)$:
\begin{eqnarray}
ds^2 &=& -\left(1 - \frac{2r}{\Sigma}M_{\text{eff}}(r)\right)dt^2
- \frac{4ar}{\Sigma}M_{\text{eff}}(r)\sin^2\theta dtd\phi
\nonumber\\
&& + \sin^2\theta
\left[r^2+a^2+\frac{2a^2r}{\Sigma}M_{\text{eff}}(r)\sin^2\theta\right]d\phi^2
\nonumber \\
&&+ \frac{\Sigma}{\Delta}dr^2+\Sigma d\theta^2.
\end{eqnarray}
Here $a$ is a spin parameter of the black hole corresponding to
the angular momentum of unit mass, and $M_{\text{eff}}(r)$ is an
effective mass  as a function of $r$:
\begin{eqnarray}
M_{\text{eff}}(r) =MG(r)= M\left(1 - \frac{\xi}{r^2}\right).
\end{eqnarray}
The other notations are
\begin{eqnarray}
\Delta = r^2 - 2Mr\left(1 - \frac{\xi}{r^2}\right) + a^2,
\end{eqnarray}
and
\begin{eqnarray}
\Sigma = r^2 + a^2\cos^{2}\theta.
\end{eqnarray}

Clearly, the metric (2) is the Kerr metric  for the quantum
parameter $\xi=0$. Unlike that in the RGI  Kerr metric, the Newton
gravitational constant in the RGI Schwarzschild black hole given
by Bonanno and Reuter [3] is modified as
 \begin{eqnarray}
 G_{Sch}(r)= \frac{G_0r^3}{r^3+\Omega G_0(r+\gamma G_0M)},
 \end{eqnarray}
where $\Omega$ and $\gamma$ are two dimensionless quantum
correction parameters. When the distance $r$ is sufficiently
large,  $G_{Sch}(r)$ can be expanded as
 \begin{eqnarray}
 G_{Sch}(r)\approx G_0 \left(1-\frac{\Omega G_0}{r^2}-\frac{M\gamma}{r^3}\Omega G^{2}_{0}+\cdots \right).
 \end{eqnarray}
By the comparison between Eqs. (1) and (7), we have
\begin{equation}
\xi=\Omega.
\end{equation}
This shows that the quantum parameter in the RGI
Kerr metric [5] is the same as one of the two quantum parameter in
the RGI Schwarzschild spacetime [3]. As was mentioned in the
Introduction Section, the running Newton constants in Equations
(1) and (6) are somewhat different. The running Newton constant
(1) in the RGI Kerr metric of Ref. [5] was given in the infra-red
limit of asymptotically safe gravity. However, the running Newton
constant (6) in the RGI Schwarzschild spacetime of Ref. [3] was
obtained from the exact evolution equation for the effective
average action. Therefore, the RGI Kerr black hole (2) in Ref. [5]
is also slightly unlike the RGI rotating black hole in Ref. [4].
Considering the measured diameter of the shadow of the M87*
central black hole, Lu and Xie [6] constrained  the parameter
$\Omega$ in the range of $0.165 \leq \Omega \leq 9.804$, that is,
\begin{equation}
0.165 \leq \xi \leq 9.804.
\end{equation}

An external asymptotically homogeneous magnetic field surrounding
the RGI  Kerr black hole is described in terms of an
electromagnetic four-vector potential of Wald [22]
\begin{eqnarray}
A^{\mu} &=& aB\xi^{\mu}_{(t)}+\frac{B}{2}\xi^{\mu}_{(\phi)},
\end{eqnarray}
where $\xi^{\mu}_{(t)}=(1,0,0,0)$ and
$\xi^{\mu}_{(\phi)}=(0,0,0,1)$ are timelike and spacelike axial
Killing vectors, and $B$ is the magnetic field strength. Two
nonzero covariant components of the four-vector potential
(see e.g. [23]) are
\begin{eqnarray}
A_{t} &=& aBg_{tt}+\frac{B}{2}g_{t\phi}, \nonumber \\
&=& -aB[1+\frac{rM_{\text{eff}}(r)}{\Sigma}(\sin^{2}\theta-2)], \\
A_{\phi} &=& aBg_{t\phi}+\frac{B}{2}g_{\phi\phi} \nonumber  \\
&=& B\sin^{2}\theta[\frac{r^2+a^2}{2}
+\frac{a^2r}{\Sigma}M_{\text{eff}}(r) \nonumber \\
&& \times(\sin^{2}\theta-2)].
\end{eqnarray}
Seen from the suggestion of Azreg-A\"{i}nou [24], the Wald
potential seems to dissatisfy the source-less Maxwell field
equations and should be generalized appropriately because the RGI
Kerr black hole is given for the case of a nonvacuum black hole in
the modified gravity. However, we use the idea of Azreg-A\"{i}nou
to easily show that the Wald potential is still a solution of the
source-less Maxwell field equations in the RGI Kerr black hole
spacetime. In fact, the function $f(r)=f_1r+f_2$ in Eq. (15) of
the Azreg-A\"{i}nou's paper is expressed in the present case as
$f(r)=rM_{\text{eff}}(r)=Mr-M\xi/r$. That is, $f_1=M$ and $f_2=0$.
In this case, Eq. (36) of the Azreg-A\"{i}nou's paper is Eq. (10)
in the present paper.

Consider that a test particle with mass $m_p$ and charge $q$ moves
around the RGI Kerr black hole surrounded by the external magnetic
field. The particle motion is given by the following Hamiltonian
\begin{eqnarray}
H = \frac{1}{2m_p} g^{\mu\nu} (p_\mu - q A_\mu) (p_\nu - q A_\nu).
\end{eqnarray}
The contravariant metric has nonzero components
\begin{eqnarray}
g^{tt}&=&-\frac{(r^2+a^2)^2-\Delta a^2\sin^{2}\theta}{\Delta\Sigma}, \\
g^{t\phi}&=&-\frac{2ar}{\Delta\Sigma}M_{\text{eff}}(r)=g^{\phi t},  \\
g^{rr}&=&\frac{\Delta}{\Sigma},\\
g^{\theta\theta}&=&\frac{1}{\Sigma},  \\
g^{\phi\phi}&=&\frac{\Sigma-2rM_{\text{eff}}(r)}{\Delta\Sigma\sin^{2}\theta}.
\end{eqnarray}
The covariant generalized four-momentum $p_{\mu}$ is determined by
\begin{eqnarray}
\dot{x}^{\mu}&=&\frac{\partial H}{\partial p_{\mu}}
=\frac{1}{m_p}g^{\mu\nu}(p_{\nu} -qA_{\nu}).
\end{eqnarray}
This equation has its equivalent form
\begin{eqnarray}
p_\mu=m_pg_{\mu \nu}\dot{x}^\nu+qA_\mu.
\end{eqnarray}
$\dot{x}^\mu=(\dot{t},\dot{r},\dot{\theta},\dot{\varphi})$ is a
four-velocity as a derivative of the coordinate $x^\mu$ with
respect to the proper time $\tau$.

Dimensionless operations are given to the Hamiltonian system (13)
through a series of scale transformations:  $r\rightarrow rM$,
$t\rightarrow tM$,$\tau\rightarrow \tau M$, $a\rightarrow aM$,
$E\rightarrow Em_p$,$p_r\rightarrow m_pp_r$, $L\rightarrow m_pML$,
$p_{\theta}\rightarrow m_pMp_{\theta}$, $q\rightarrow m_pq$,
$B\rightarrow B/M$ and $H\rightarrow m_pH$. Thus, $m_p$ and $M$ in
Eqs. (1)-(20) become two mass factors eliminated. Of course, these
scale transformations describe the correspondence of the
dimensionless quantities and the practical ones. For instance, the
dimensionless angular momentum $L$ corresponds to the practical
one $m_pML$; when the dimensionless magnetic field strength is
$B$, the practical one is $B/M$. The strength of the magnetic
field in the SI units is given in [15] by
\begin{eqnarray}
B_{SI}=\frac{\beta c_{SI}}{q_{SI}1472m(M/M_{\odot})}, \nonumber
\end{eqnarray}
where $\beta=qB$, the specific charge of an electron is
$q_{SI}=1.76\times 10^{11}C/(kg)$ and $c_{SI}$ is the speed of
light in the SI units. The length 1472 m represents the solar mass
in geometrized units, $M_{\odot}= 1472m$. In this way, the
dimensionless quantities can be mapped to typical astrophysical
scales. For example, the value of magnetic field strength near a
black hole with mass $M=10M_{\odot}$ and $\beta=1$ is
$B_{SI}=1.16\times 10^{-7}T$.

The dimensionless Hamiltonian (13) does not explicitly depend on
the coordinates $t$ and $\phi$. Therefore, $p_t$ and $p_\phi$ as
constants of motion correspond to the particle's energy $E$ and
angular momentum $L$:
\begin{eqnarray}
p_t &=&g_{tt}\dot{t}+g_{t\phi}\dot{\phi} +qA_{t}=-E,\\
p_\phi &=& g_{t\phi}\dot{t}+g_{\phi\phi}\dot{\phi} +qA_{\phi}=L.
\end{eqnarray}
Now, the dimensionless Hamiltonian (13) has two degrees of freedom
$(r,\theta)$ and its phase space has four dimensions. In this
case, the Hamiltonian is rewritten as
\begin{eqnarray}
H = F +\frac{1}{2}\frac{\Delta}{\Sigma}p^{2}_{r}
+\frac{1}{2}\frac{p^{2}_{\theta}}{\Sigma},
\end{eqnarray}
where $F$ is a sub-Hamiltonian in the form
\begin{eqnarray}
F &=&\frac{1}{2}[g^{tt}(E+qA_{t})^2 +g^{\phi\phi}(L-qA_{\phi})^2] \nonumber \\
&& -g^{t\phi}(E+qA_{t})(L-qA_{\phi}).
\end{eqnarray}
Due to the rest mass for a time-like geodesic, the dimensionless
Hamiltonian (23) has a third constant of motion
\begin{eqnarray}
    H=-\frac{1}{2}.
\end{eqnarray}

Given $\beta=0$, the system (23) possesses the Carter constant as
a fourth motion constant and therefore is integrable. However, the
Carter constant is absent and the system (23) becomes
nonintegrable if $\beta\neq0$.

\section{Numerical investigations}

At first, explicit symplectic algorithms are designed  for the
Hamiltonian system (23). Then, one of them is applied to
investigate the dynamics of the Hamiltonian system.

\subsection{Design of explicit symplectic integrators}

The Hamiltonian system (23) cannot be directly split into several
explicit integrable pieces. However, it can through an appropriate
time transformation suggested in Ref. [25]. Set $\tau $ as a new
coordinate $q_0=\tau$, and its corresponding momentum is
$p_0=-H=1/2$. The phase space variables $(p_r,p_\theta,r,\theta )$
of the Hamiltonian (23) are extended to the phase space variables
$(p_r,p_\theta,p_0,$ $ r,\theta,q_0)$. An extended phase-space
Hamiltonian is written as
\begin{eqnarray}
     \mathcal{H} =H+p_0.
\end{eqnarray}
It is clear that $ \mathcal{H}$ is always identical to zero, $
\mathcal{H}=0$.

Taking a time transformation
\begin{eqnarray}
    d\tau &=& g(r,\theta)dw,
\end{eqnarray}
where $g(r,\theta)$  is a time transformation function
\begin{eqnarray}
   g(r,\theta)=\frac{\Sigma}{\Delta},
\end{eqnarray}
we obtain a new time transformation Hamiltonian
\begin{eqnarray}
J=g \mathcal{H}&=&\frac{\Sigma}{\Delta} (F+p_0)
+\frac{1}{2}p^{2}_{r} +\frac{1}{2\Delta}p^{2}_{\theta}.
\end{eqnarray}
The time-transformed Hamiltonian has the following split
\begin{eqnarray}
   J=J_1+J_2+J_3,
\end{eqnarray}
where $J_1$, $J_2$ and $J_3$ are three sub-Hamiltonian systems:
\begin{eqnarray}
    J_1&=&\frac{\Sigma}{\Delta} (F+p_0),\\
    J_2&=&\frac{p_{r}^2}{2},\\
    J_3&=&\frac{p_\theta^2}{2 \Delta}.
\end{eqnarray}
The three splitting parts have analytical solutions, which are
explicit functions of the proper $\tau$. The solvers of these
sub-Hamiltonians are labeled as $\mathcal{J}_1$, $\mathcal{J}_2$
and $\mathcal{J}_3$  in turn. The computational
efficiency of such a three-part splitting form in the RGI Kerr
case is relatively superior to that of the six-part splitting form
in the RGI Schwarzschild one [13]. 

Let $h$ be a time step of the new time $w$. Two first-order
symplectic composing operators are
\begin{eqnarray}
    \chi (h)&=&  \mathcal{J}_3(h)\times\mathcal{J}_2(h)\times\mathcal{J}_1(h), \\
    \chi^{*}(h) &=&  \mathcal{J}_1(h)\times\mathcal{J}_2(h)\times\mathcal{J}_3(h).
\end{eqnarray}
In terms of the two operators, an explicit second-order symplectic
method is given by
\begin{eqnarray}
    S2(h)=\chi(\frac{h}{2})\times\chi^{*}(\frac{h}{2}).
\end{eqnarray}
On the basis of the second-order method, an explicit fourth-order
symplectic scheme of Yoshida [26] can be produced by
\begin{eqnarray}
    S4(h)=S2(\sigma h)\times S2(\delta h)\times S2(\sigma h),
\end{eqnarray}
where $\delta=1-2\sigma$ and $\sigma=1/(2-\sqrt[3]{2})$.

The time step is taken as $h=1$. The parameters are chosen as
$E=0.995$, $L=4.6$,  $\beta=1\times 10^{-3}$, $a = 0.5$ and $\xi =
0.2$. Two orbits have the initial conditions  $p_{r}=0$ and $
\theta =\pi/2$. The initial radius is $r=30$ for Orbit 1. It is
$r=80$ for Orbit 2. The positive initial values of $p_{\theta}$
for the two orbits are solved from Eq. (25). When the integration
time $w$ reaches $10^7$, the two algorithms S2 and S4 give no
secular growth to the Hamiltonian errors $\Delta H=H+1/2$ (i.e.
$\mathcal{H}$) for the two orbits in Fig. 1 (a) and (b). This
result is what symplectic integrators own. The error of S4 is
three orders of magnitude smaller than that of S2 for each of the
two orbits. Because the accuracy of the method S4 is greatly
superior to that of the method S2, it is used to simulate the
dynamics of charged particles.

Although the time transformation (27) is considered, there is no
typical difference between the new time $w$ and the original time
$\tau$, as shown  in Fig. 1(c) about S4 describing the relation
between $w$ and $\tau$. This fact is seen clearly from the time
transformation function $g(r,\theta)\rightarrow 1$ for a
sufficiently large distance $r$ in Eq. (28). In other words,
neither S4 nor S2 adopts adaptive time steps for the original time
$\tau$ when a constant time step is employed for the new time $w$.
More recently, adaptive time step explicit symplectic methods were
designed for many curved spacetimes in Ref. [27]. Another notable
point is that the new time transformation function and the
three-part split of the time-transformed Hamiltonian (30) bring
better accuracy and computational efficiency than the original
five-part split [24] or the original six-part split [13].

\subsection{Orbital dynamics}

Using the algorithm S4 and several chaos indicators, we provide an
insight into some differences in the dynamics of Orbits 1 and 2.
Then, we mainly focus on how the parameter $\xi$ makes a
contribution to chaos.

\subsubsection{Dynamical features of Orbits 1 and 2}

The Poincar\'{e} map in the two-dimensional $r-p_r$ plane shows
the regularity of Orbit 1 in Fig. 2(a) because  intersections of
Orbit 1 with the surface of section in phase space form a torus.
However, the random distribution of the points in the
Poincar\'{e} map describes the characteristic of chaotic behavior
on Orbit 2. Seen from Figs. 1(a), (b) and 2(a), the accuracy of
the Hamiltonian $\mathcal{H}$ has no close link with chaotic or
regular motion of a charged particle near the black hole. On the
contrary,  the error of the solution $(r,\theta,p_r,p_{\theta})$
is drastically affected by the chaotic or regular dynamics of a
charged particle. It would increase exponentially with time, in
particular for the case of chaos.  This attributes to the basic
characteristic of chaos, which makes the error exhibit sensitive
dependence on the initial errors.

Apart from the numerical error of the solution having the
sensitivity to the initial errors, the solution
$(r,\theta,p_r,p_{\theta})$ has the sensitivity to the initial
values. The sensitivity is quantitively evaluated in terms of
Lyapunov exponents. The largest Lyapunov exponent in [28] is
calculated by
\begin{eqnarray}
\lambda=\lim_{w\rightarrow\infty}\frac{1}{w}\ln\frac{d(w)}{d(0)},
\end{eqnarray}
where $d(w)$ and $d(0)$ stand for the proper distances of two
nearby orbits at the proper time $w$ and the starting time,
respectively. When the two nearby orbits are bounded, a positive
value of the Lyapunov exponent means that the difference between
the two nearby orbits grows exponentially. In this way, chaotic
dynamics is shown. See also [29] for more details on the Lyapunov
characteristic exponents and their computation. The vanishing
Lyapunov exponent describes regular dynamics because the
difference oscillates within a certain range or grows in a power
law. In Fig. 2(b), $\lambda\rightarrow$ a positive value shows the
chaoticity of Orbit 2, whereas $\lambda\rightarrow 0$ indicates
the regularity of Orbit 1.

In fact, a long enough integration time is necessary in general
for $\lambda\rightarrow$ the positive value or $\lambda\rightarrow
0$. A quicker method for finding chaos is fast Lyapunov indicators
(FLIs) [30]. The Lyapunov exponent of two nearby orbits
corresponds to the FLI of two nearby orbits [28]:
\begin{eqnarray}
    FLI=\log _{10}\frac{d(w)}{d(0)}.
\end{eqnarray}
An exponential increase of the FLI with time $\log _{10}w$ means
the exponential deviation of two adjacent bounded orbits and shows
the presence of chaos. On the other hand, a slow increase of the
FLI with time $\log _{10}w$ means the algebraic deviation of two
adjacent bounded orbits and shows the presence of order. In this
way, the regular dynamics of Orbit 1 and the chaotic dynamics of
Orbit 2 are clearly confirmed by the FLIs in Fig. 2(c).

In addition to the above methods, the 0-1 test is a reliable
method for chaos [31]. In this method, chaotic behavior is
determined through the growth rate of the mean square displacement
$M_c(n)$ of two translation variables $p_c(n)$ and $q_c(n)$ as a
function of time $n$, where $c\in(0,\pi)$ is a free parameter. The
function in [28] is of the expression $M_c(n)= V_1(c)n+
V_2(c,n)+e(c,n)$, where $V_1\geq0$ is a finite number, $V_2$ is a
bounded oscillatory term and $e(c,n)/n\rightarrow0$ as
$n\rightarrow\infty$. Note that $V_1=0$ for  periodic or
quasiperiodic motion but $V_1>0$ for chaotic motion. As a result,
the asymptotic growth rate of the mean square displacement
$\tilde{K}_{c}=\log M_c(n)/\log n$ is 0 for the former case but 1
for the latter case when $n\rightarrow\infty$. The 0-1 test for
chaos is the regression method, which consists of linear
regression for the log-log plot of the mean square displacement
[31]. As a refined version of the regression method with
$\tilde{K}_{c}$, the correlation method [32] is a sensitive 0-1
test for chaos. In this case, the correlation coefficient $K_c=0$
or  $K_c=1$ is used to signify nonchaotic or chaotic dynamics. In
Fig. 3, Orbit 1 with $K_c=0.04627\approx 0$ is regular and Orbit 2
with $K_c=0.98774\approx 1$ is chaotic. The visual plot on the
translation components $(p_c, q_c)$ is a bounded region, which
corresponds to the regular dynamics of Orbit 1. An asymptotic
unbounded Brownian motion as the visual plot of the translation
components $(p_c, q_c)$ also shows the chaotic dynamics of Orbit
2.

There are some differences among the above four techniques for
identifying chaos. The method of Poincar\'{e} map can intuitively,
clearly depict the phase space structure of a conservative
Hamiltonian system with two degrees of freedom. The methods of
Lyapunov exponents, FLIs and 0-1 test are suitable for any
dimensional systems. The method of FLIs is a quicker tool to
distinguish between order and chaos than that of Lyapunov
exponents. Both the FLI  and the 0-1 test are convenient to trace
a transition from regular dynamics to chaotic dynamics when one or
two parameters vary. The FLI greatly outperforms the 0-1 test in
computational efficiency.

\subsubsection{Effect of the parameter $\xi$ on chaos}

Now, let us take the starting separation $r=30$ and the parameters
$E=0.995$, $L=4.5$, $\beta=8\times 10^{-4}$ and $a=0.67$. The
parameter $\xi$ is given three values: $\xi=0.18$, $\xi=5.56$ and
$\xi=8.56$. The Poincar\'{e} map in Fig. 4(a), the correlation
method of 0-1 test in Fig. 4(d) and the techniques of  Lyapunov
exponents and FLIs in Fig. 5(a) and (b) claim together that the
dynamics is chaotic for $\xi=0.18$. Nevertheless,  the regular
dynamics are demonstrated for $\xi=5.56$ and $\xi=8.56$ in Figs. 4
(b), (c), (e), (f) and 5 (a), (b). A result concluded from these
facts seems to be that the extent of chaos is weakened or
suppressed as the parameter $\xi$ increases. To use more examples
to support this point, we plot the dependence of the FLI on $\xi$
ranging from 0 to 9.8 in Fig. 5(c). Each of the FLIs is not
obtained until the integration time is $w=5\times 10^6$. It is
found that the transition from chaotic dynamics to regular
dynamics begins to occur as the parameter spans $\xi=0.65$, where
FLI=5.87 is considered as a threshold between order and chaos.

Taking the initial separation $r=70$, we use FLIs to find chaos in
two parameter spaces in Fig. 6. Chaos is easier for smaller
angular momentum $L$, span $a$ or quantum corrected parameter
$\xi$. It is also for larger magnetic parameter $\beta$ and energy
$E$. In particular, it is also shown through the four methods and
various combinations of other parameters in Figs. 7-10 that a
decrease of the quantum corrected parameter $\xi$ is possible to
easily induce chaos. Of course, these results are considered from
a statistical point of view. It is not always correct that a
smaller value of the quantum corrected parameter must bring
stronger chaos than  a larger value of the quantum corrected
parameter when the other parameters and the initial conditions
(excluding the initial value of $p_{\theta}$) are given. However,
it should be true that the probability for the occurrence of chaos
or strong chaos is larger in the case of small values of the
quantum corrected parameter than that  in the case of large values
of the quantum corrected parameter. The increase
of the quantum improved parameter $\xi$ weakening the extent of
chaos for the case of dragging effects of the spacetime is similar
to that of the quantum improved parameter $\Omega$ in the RGI
Schwarzschild case [13].

Here, some theoretical interpretations are given to the
above-mentioned numerical results regarding the effects of the
parameters such as $\xi$ on chaos. Considering the case of
$0\leq\beta\ll1$ and $r\gg1$, we expand the gravitational
potential function $F$ of Eq. (24)  as
\begin{eqnarray}
F &\approx& \frac{\beta^2}{8}r^2+c(E, L, a, \beta,
\xi)-\frac{E^2}{r}+\frac{L^2}{2r^2} \nonumber \\
 & &+\frac{2}{r^3}aEL+\frac{\xi}{r^3}E^2+\cdots,
\end{eqnarray}
where $c(E, L, a, \beta, \xi)$ is a constant depending on $E$,
$L$, $a$, $\beta$ and $\xi$. The first term plays a role in the
Lorentz force as an attractive force, and the third term is also
an attractive force from the black hole. The fourth, fifth and
sixth terms bring a contribution of repulsive forces. In most
cases, the attractive forces are helpful to induce chaos, but the
repulsive forces are not. In fact, $\xi/r^3$ acting on the
gravitational potential from the running Newton coupling constant
$G(r)$  effectively weakens the central gravitational attraction
and reduces the steepness of the potential near the origin. This
leads to suppressing tidal forces, reducing orbital instability
and decreasing sensitivity to initial conditions. Consequently,
the extent of chaos is weakened from the statistical situation
with an increase of $\xi$. The effect of the
parameter $\xi$ on chaotic dynamics in the RGI Kerr spacetime
resembling that of the parameter $\Omega$ in the RGI Schwarzschild
case [13] can be shown clearly via comparison of Eqs. (8) and (40)
with Eq. (31) of [13].

The result is similarly suitable for an increase of $L$ or $a$,
too. However, chaos gets easier as $E$ and $\beta$ increase. On
the other hand, the magnitude of $1/r^3$ gets relatively small for
$r\gg1$ and therefore the fifth and sixth terms in Eq. (40) are
also small. This shows that the transition from chaos to order is
not highly sensitive dependence on a small change of the parameter
$\xi$ or $a$. In a word, the extent of chaos statistically, slowly
decreases with increasing $\xi$ or $a$.

\section{Summary}
\label{sec1}

By modifying the Newton gravitational constant as a function of
the radial  distance in the theory of quantum gravity, the Kerr
metric becomes the renormalization group improved (RGI) Kerr
metric. There is a quantum correction parameter in the RGI Kerr
metric. This metric is static, axial-symmetric and integrable.

When the RGI Kerr black hole is immersed in  an external
asymptotically homogeneous magnetic field, the motion of charged
particles around the black hole is nonintegrable. A new time
transformation function is given to the Hamiltonian for the
description of the motion of charged particles, and a three-part
split is applied to the time-transformed Hamiltonian. In this way,
explicit symplectic integrators are established and perform good
long-term performance.

Using  one of the explicit symplectic algorithms and chaos
indicators such as the techniques of Poincar\'{e} sections,
maximal Lyapunov exponents, fast Lyapunov indicators and  0-1
test, we find the onset of chaos under some circumstances. The
transition from regular dynamics to chaotic dynamics is traced as
one or two dynamical parameters vary. In particular, the
probability for the occurrence of chaos or strong chaos is larger
for small values of the quantum corrected parameter than that  for
large values of the quantum corrected parameter. This is because
the quantum corrected parameter is responsible for the repulsive
force which weakens the extent of chaos. Precisely speaking, the
running Newton gravity constant effectively weakens the central
gravitational attraction. This leads to reducing the steepness of
the potential near the origin, suppressing tidal forces, reducing
orbital instability and decreasing  sensitivity to initial
conditions.

\textbf{Acknowledgements}: The authors are very grateful to a
referee for valuable comments and suggestions. This research was
supported by the National Natural Science Foundation of China
(Grant No. 11973020).

\textbf{Data Availability Statement}: This manuscript has no
associated data or the data will not be deposited. [Author's
comment: All of the data are shown as the figures and formula. No
other associated data.]

\textbf{Code Availability Statement} Code/software will be made
available on reasonable request. [Author's comment: The
code/software generated during and/or analysed during the current
study is available from the first author J. Lu on reasonable
request.] All the codes are also provided in
https://github.com/LuJunjie6/Explicit-Symplectic-Integrator-and-Chaos-Indicators

\begin{figure*}[htpb]
        \centering{
        \includegraphics[width=12pc]{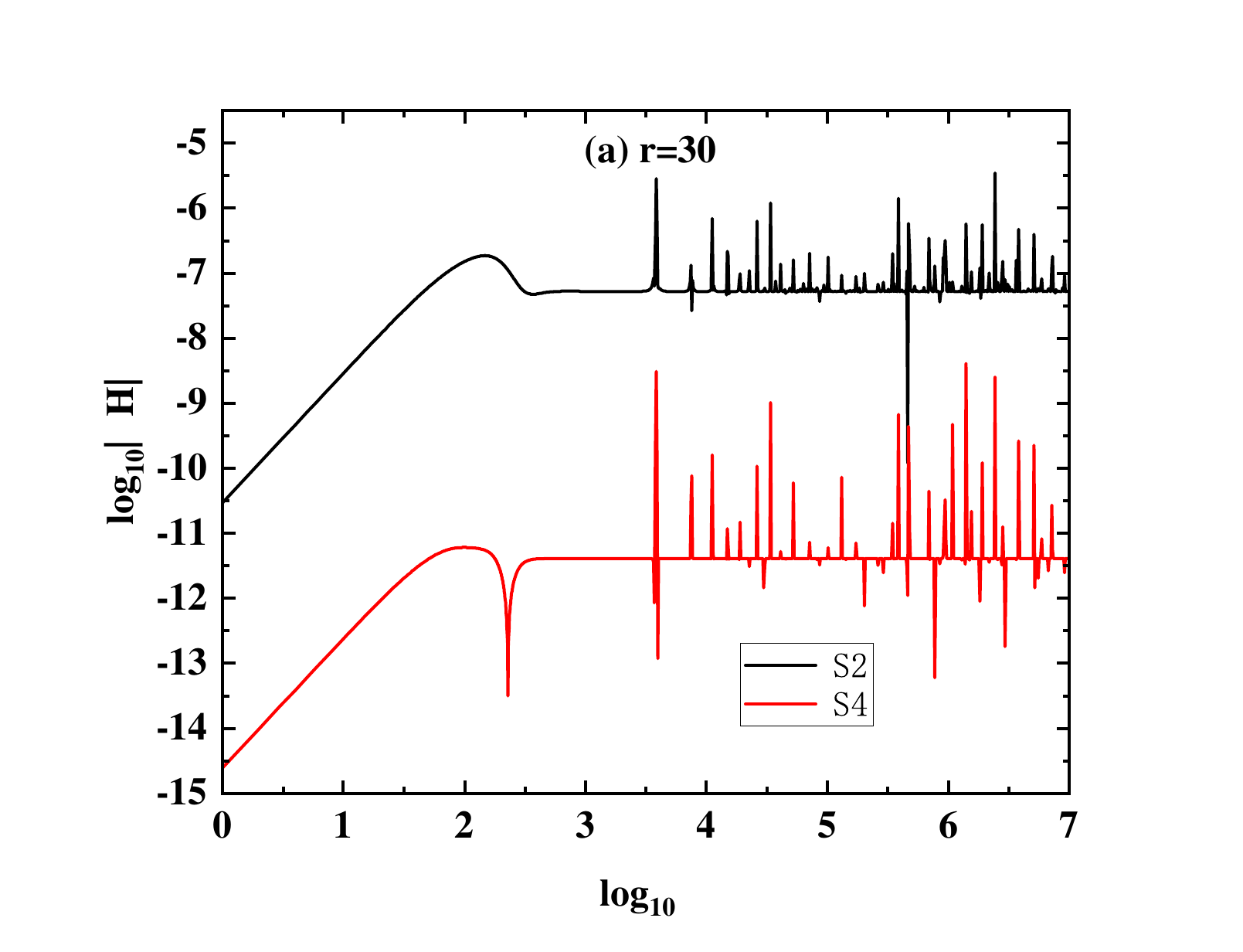}
        \includegraphics[width=12pc]{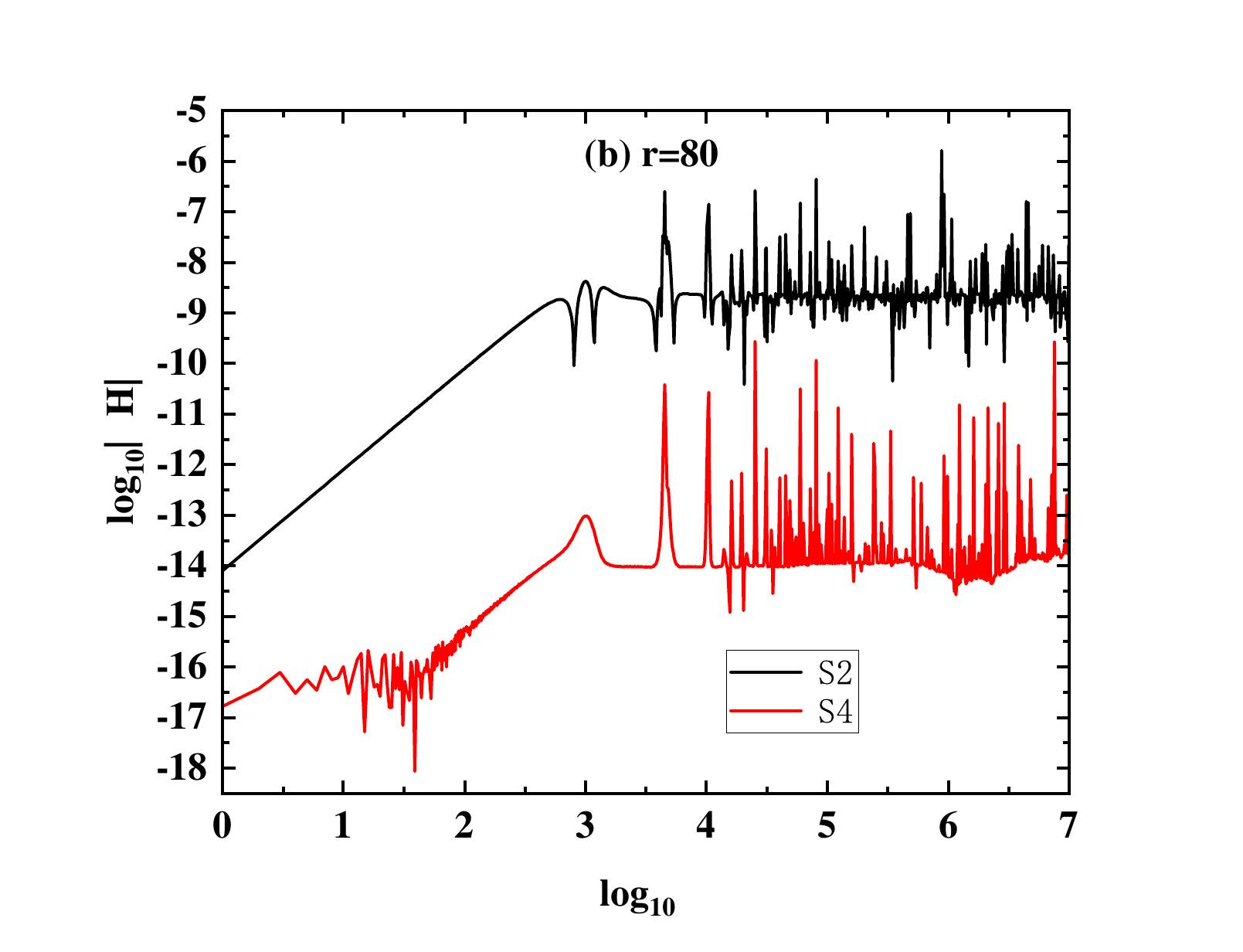}
        \includegraphics[width=12pc]{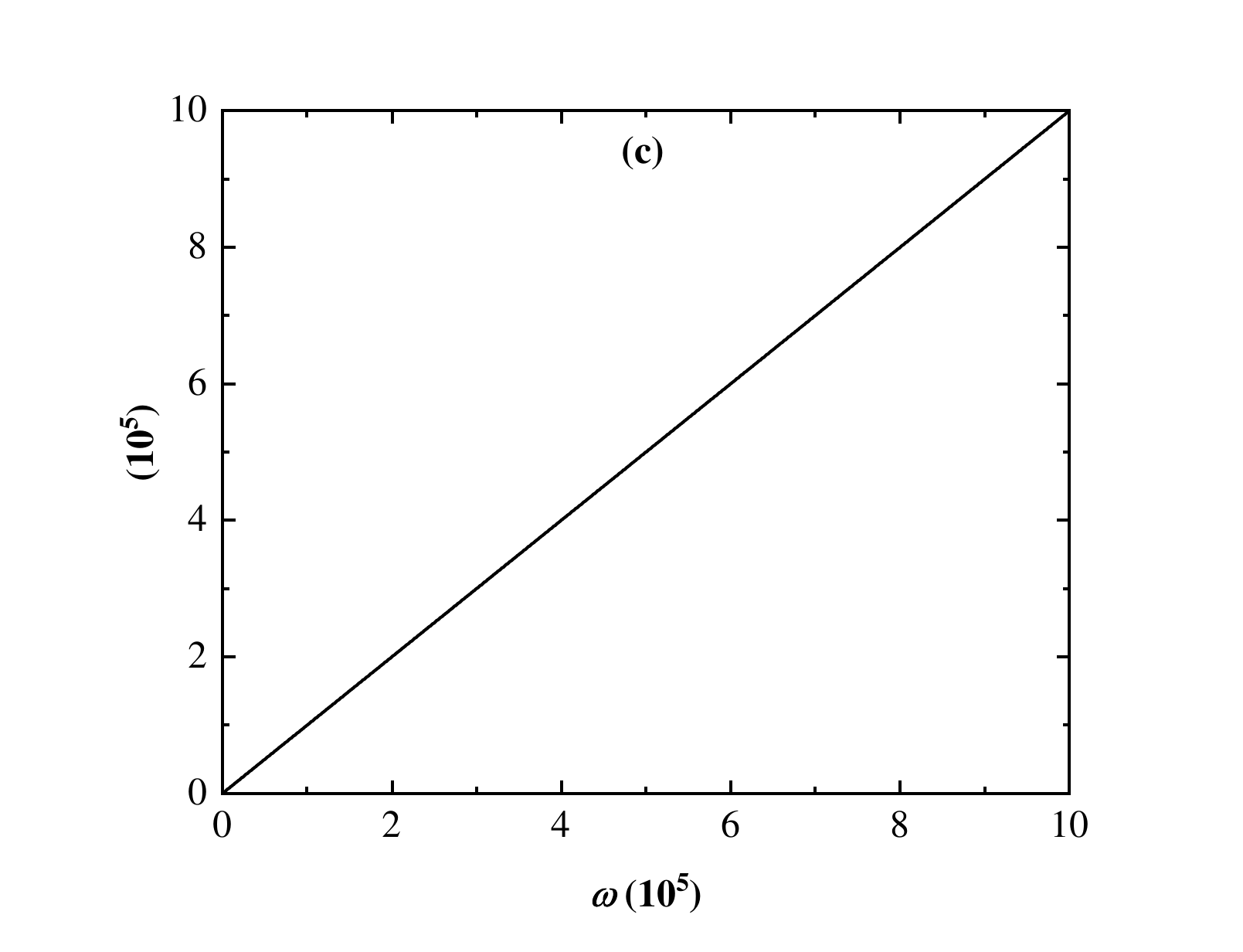}
        \caption{(a) Hamiltonian errors $\Delta H= \mathcal{H}$ for the two explicit symplectic algorithms acting on
Orbit 1 with the initial separation $r=30$. (b) Same as (a) but
Orbit 1 is replaced by Orbit 2 with the initial separation $r=80$.
Note that the two orbits correspond to the other initial
conditions $\theta=\pi/2$ and $p_{r}=0$ and the parameters
$E=0.995$, $L=4.6$,  $\beta=1\times 10^{-3}$, $a = 0.5$ and $\xi =
0.2$.  (c) The approximately equal relation between the proper
time $\tau$ and the new time $w$, described by the method S4
solving Orbit 1 or 2.
        }
    }
\end{figure*}

\begin{figure*}[htpb]
    \centering{
        \includegraphics[width=12pc]{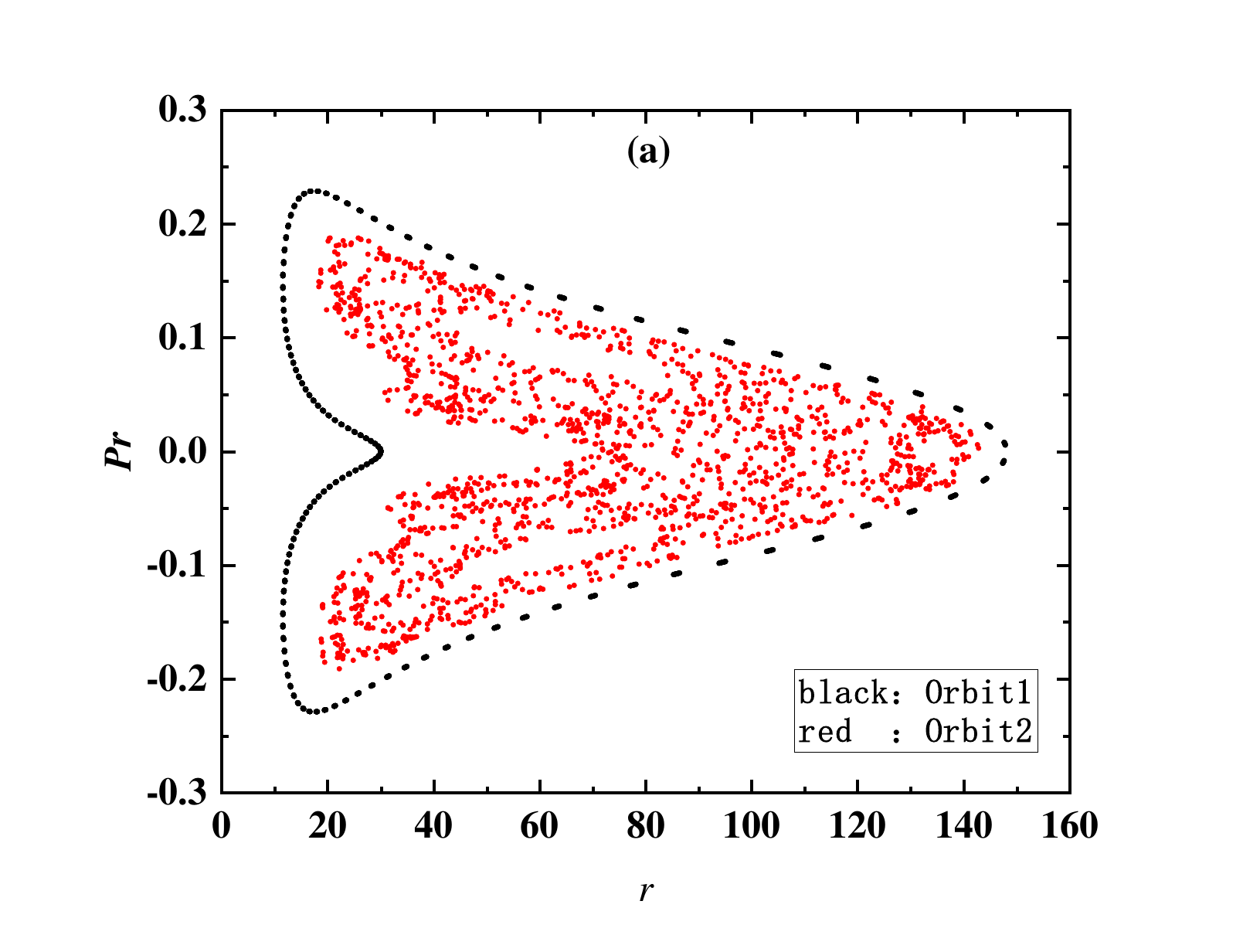}
        \includegraphics[width=12pc]{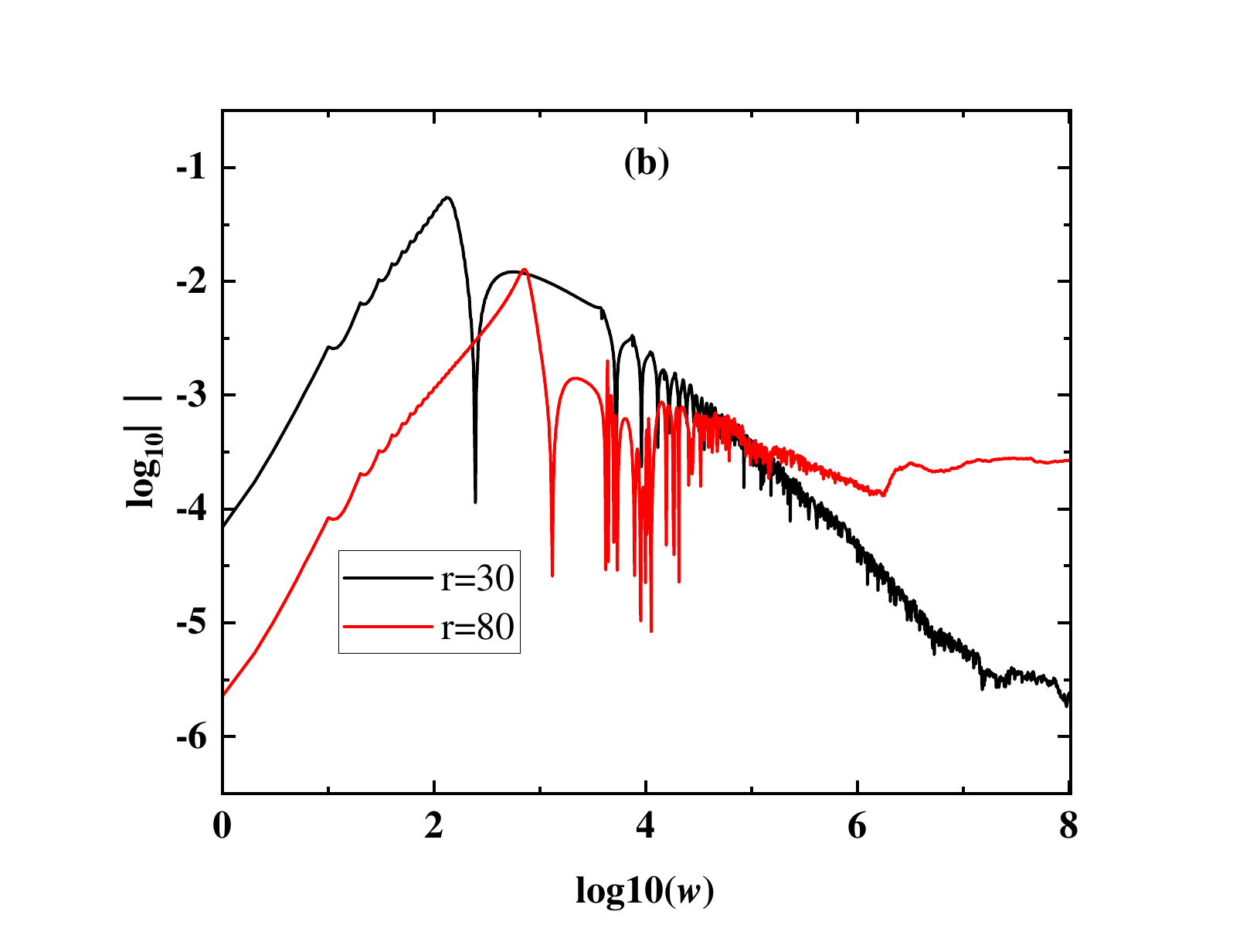}
        \includegraphics[width=12pc]{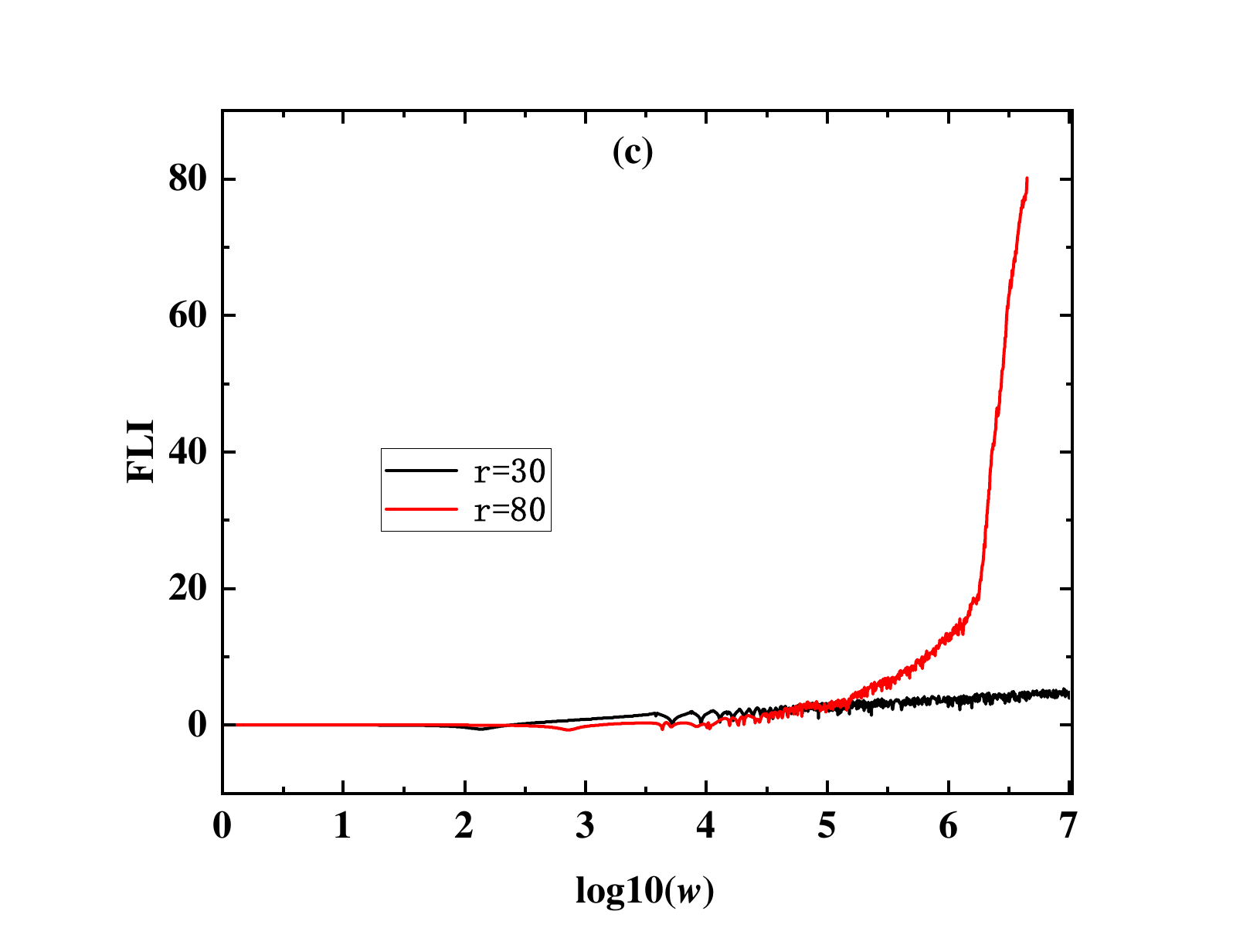}
        \caption{(a) Poincar\'{e} sections of the two orbits on the plane $\theta=\pi/2$ with $p_{\theta}>0$.
        (b) The maximal Lyapunov exponents for the two orbits.
        (c) Fast Lyapunov Indicators (FLIs) for the two orbits. The three methods consistently show
        regular dynamics of Orbit 1 and chaotic dynamics of Orbit 2.
            }
    }
\end{figure*}

\begin{figure*}[htpb]
        \centering{
        \includegraphics[width=17pc]{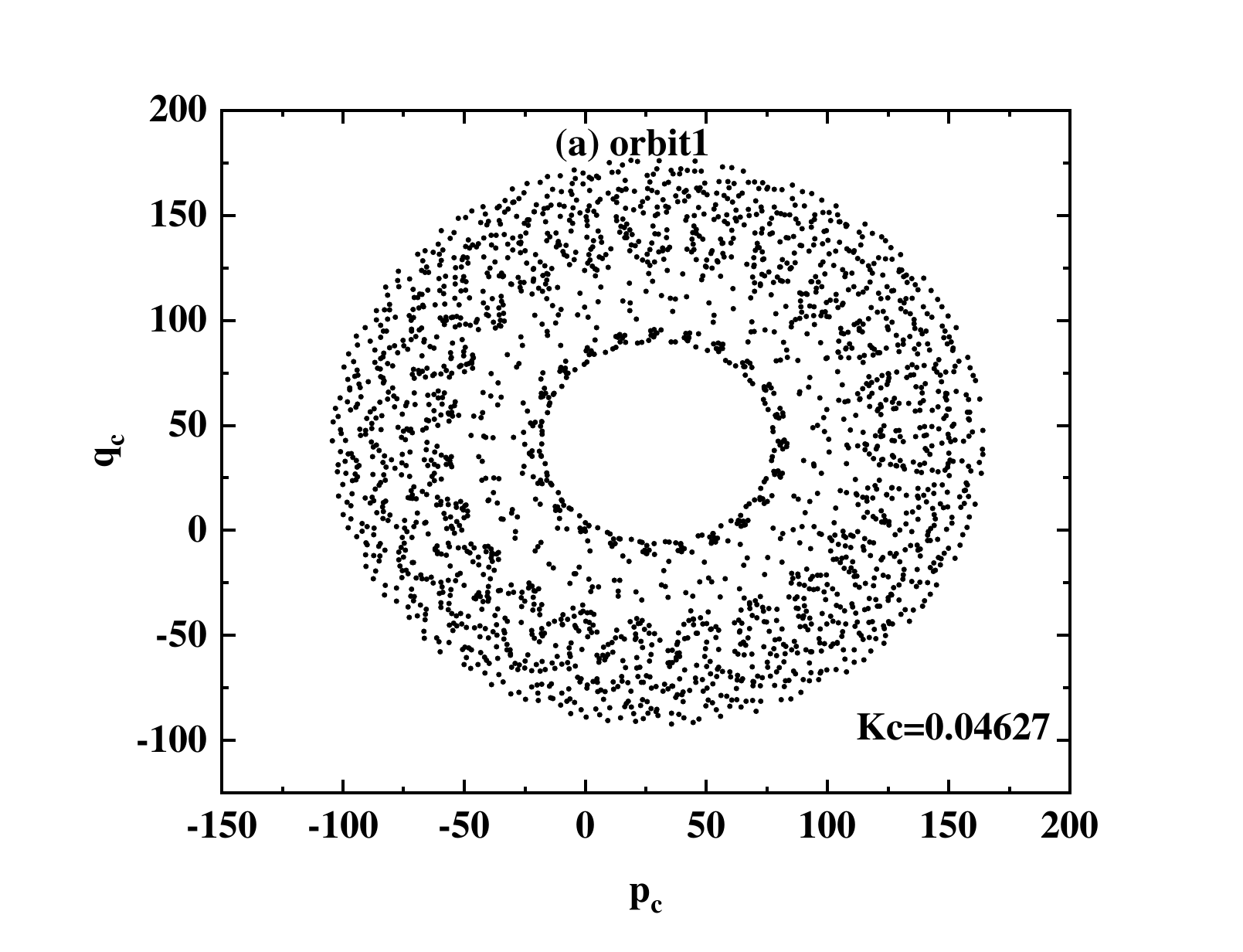}
        \includegraphics[width=17pc]{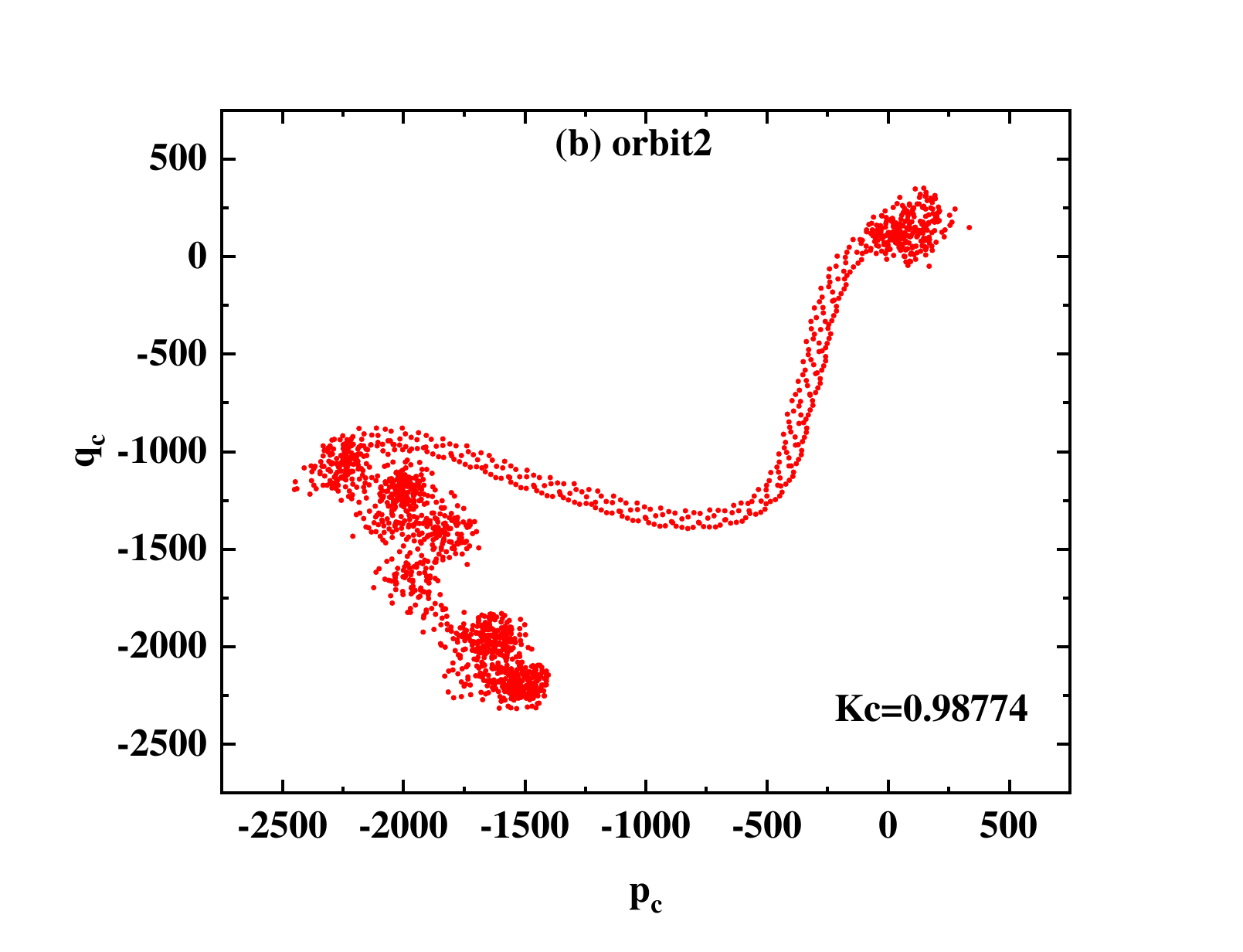}
        \caption{(a) The visual of $(p_c, q_c)$ and the correlation coefficient $K_c$ for Orbit 1. (b) Same as (a) but
        Orbit 1 is replaced by Orbit 2. The regularity of Orbit 1 and the chaoticity of Orbit 2
        are shown via the visuals $(p_c, q_c)$ and the correlation coefficients $K_c$.
        }
    }
\end{figure*}

\begin{figure*}[htpb]
    \centering{
        \includegraphics[width=12pc]{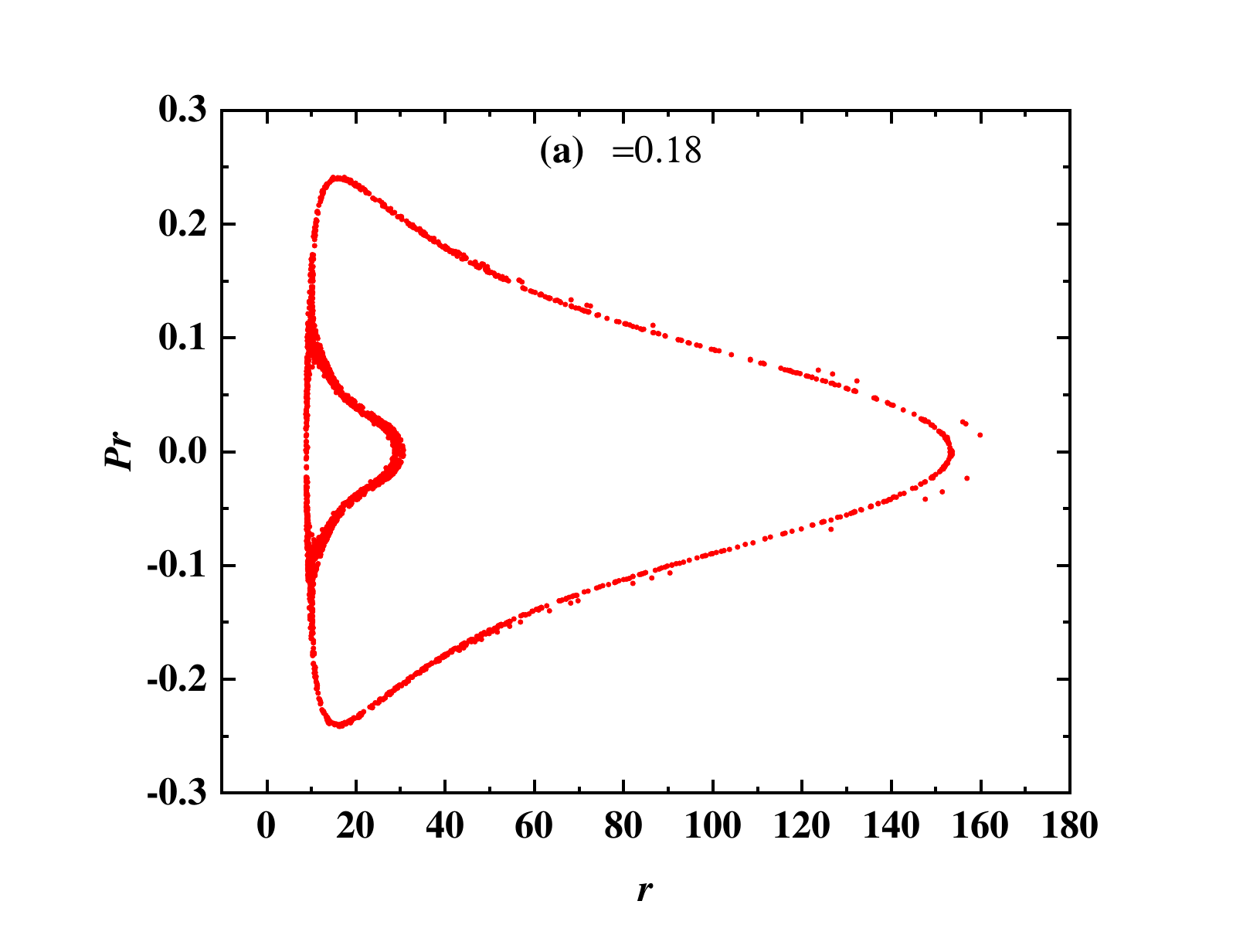}
        \includegraphics[width=12pc]{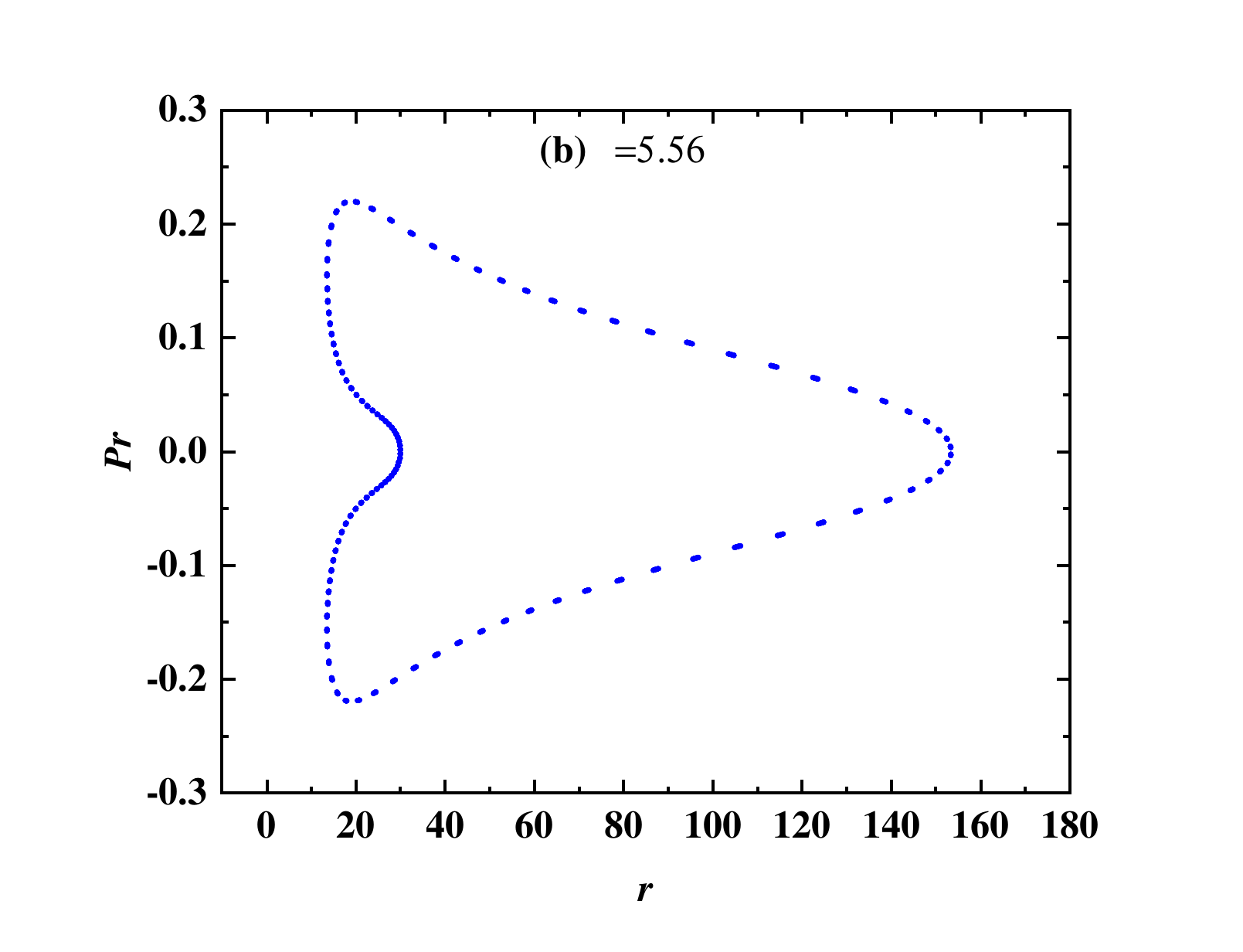}
        \includegraphics[width=12pc]{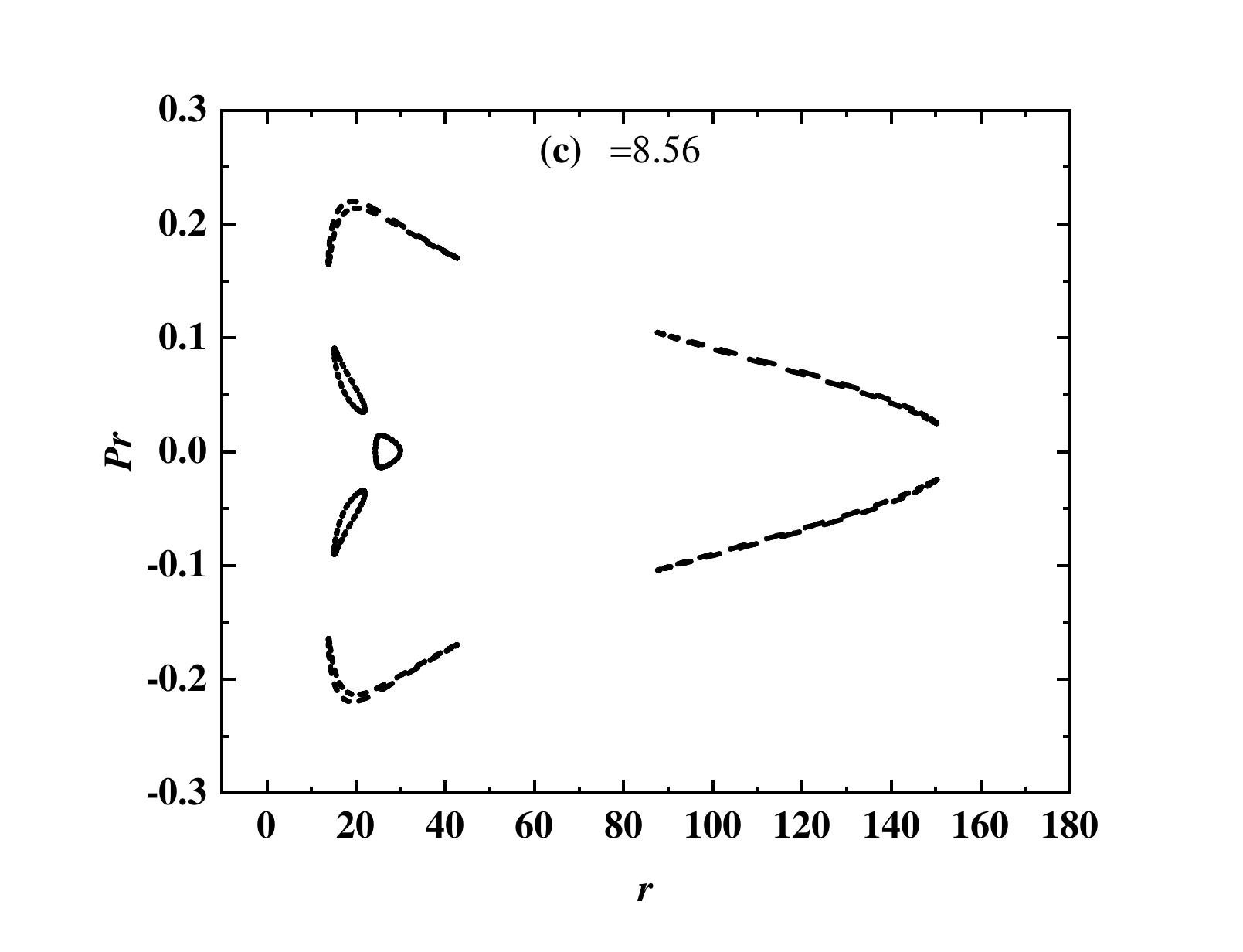}
        \includegraphics[width=12pc]{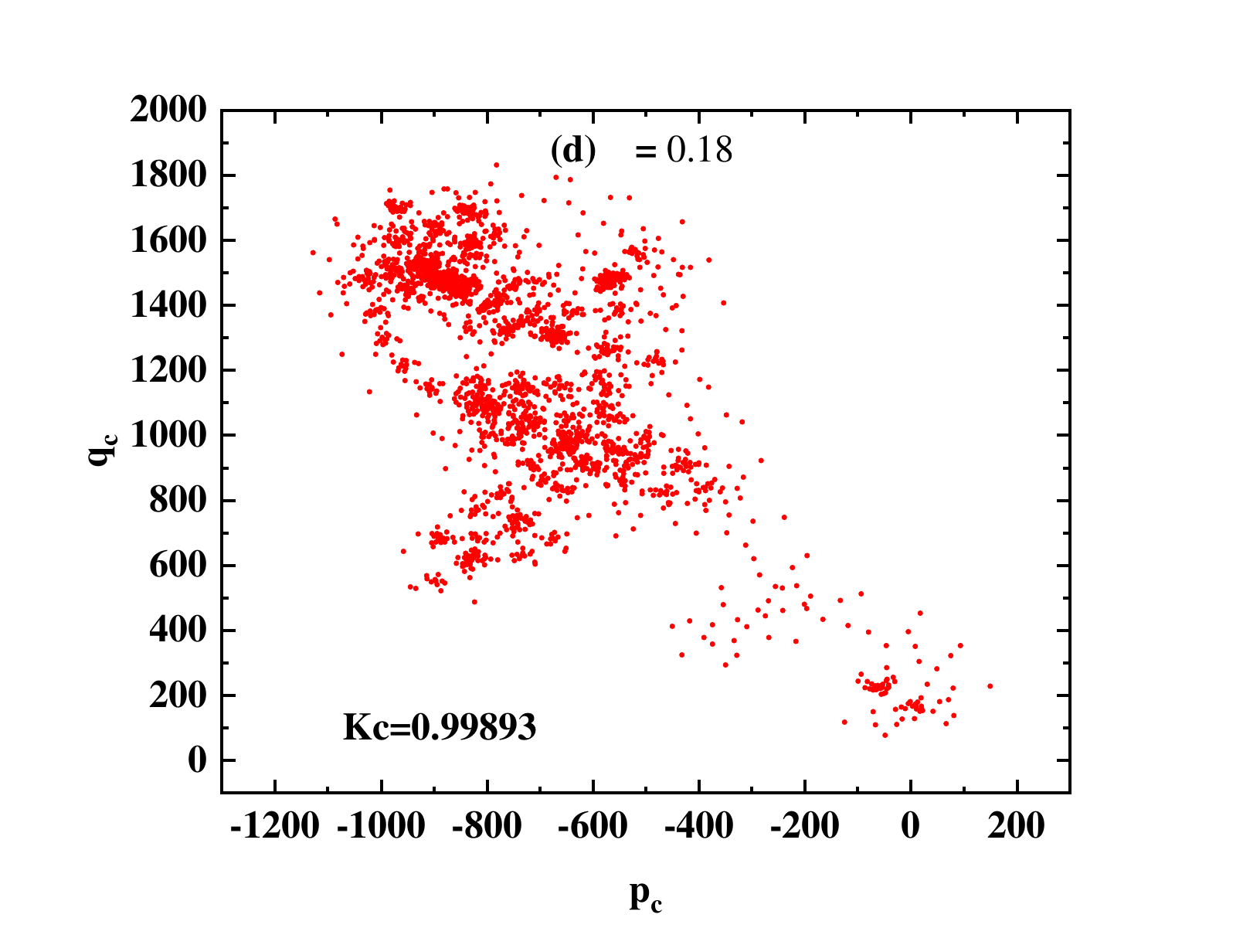}
        \includegraphics[width=12pc]{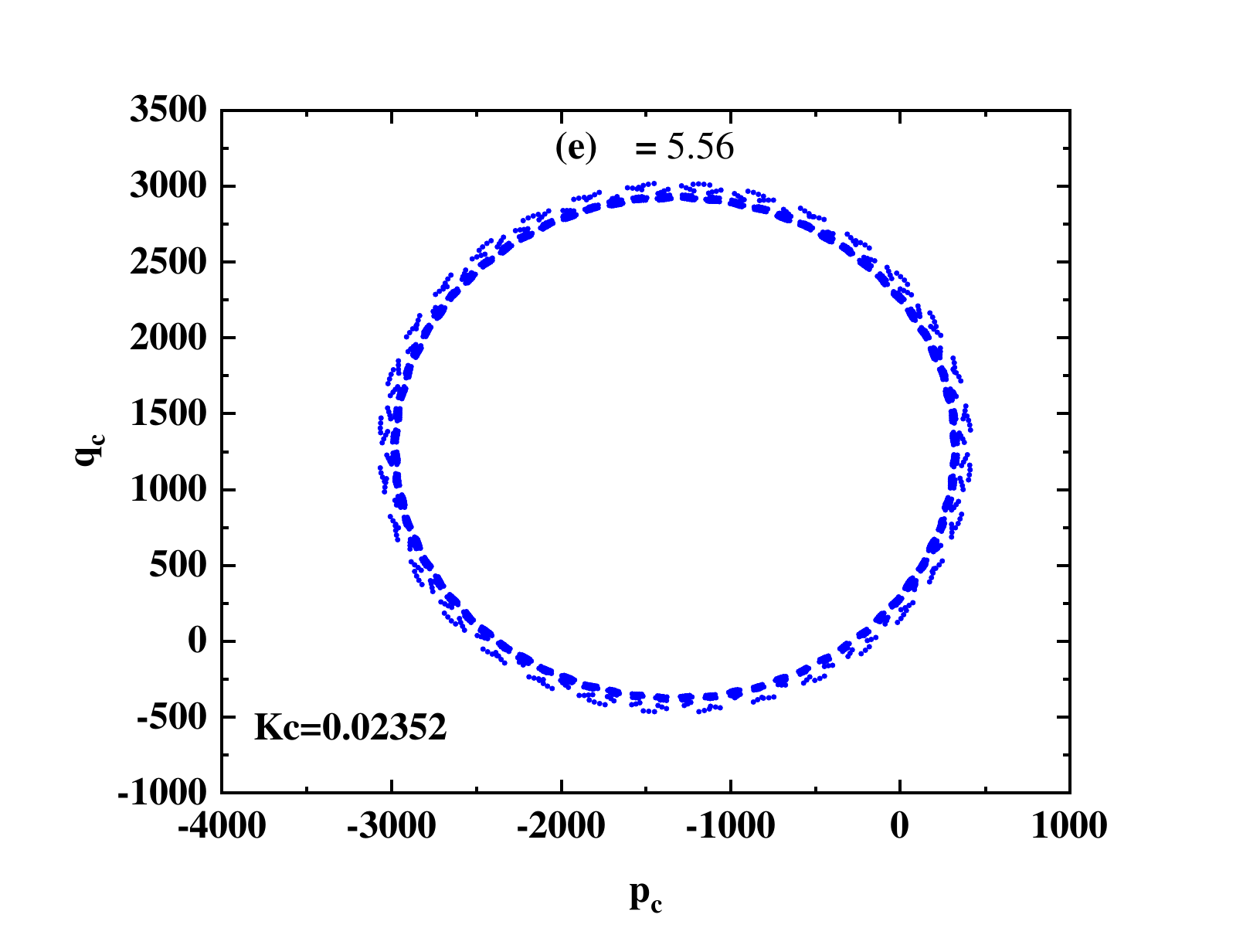}
        \includegraphics[width=12pc]{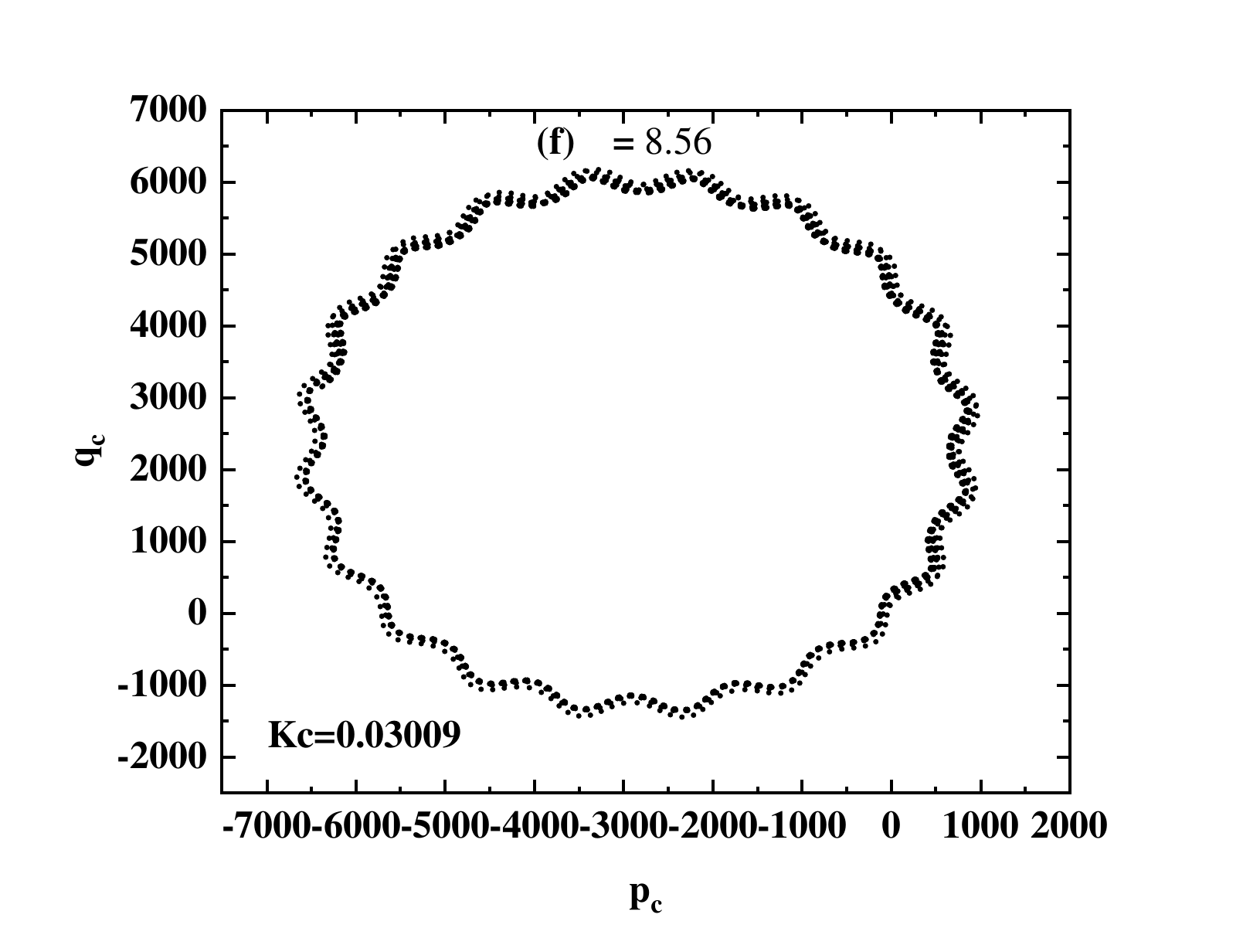}
\caption{(a)-(c): Poincar\'{e} sections describing the dynamical
behaviors of orbits for three values of the parameter $\xi$. The
other parameters are $E=0.995$, $L=4.5$, $\beta=8\times 10^{-4}$
and $a=0.67$ and the tested orbit has its initial separation
$r=30$. (d)-(f): The visuals of $(p_c, q_c)$ and the values of
$K_{c}$, which respectively correspond to panels (a-c). Chaos
occurs for $\xi=0.18$ in panels (a) and (d). Regular dynamics
exist for $\xi=5.56$ and $\xi=8.56$ in panels (b), (c), (e) and
(f).
        }
    }
\end{figure*}

\begin{figure*}[htpb]
    \centering{
        \includegraphics[width=12pc]{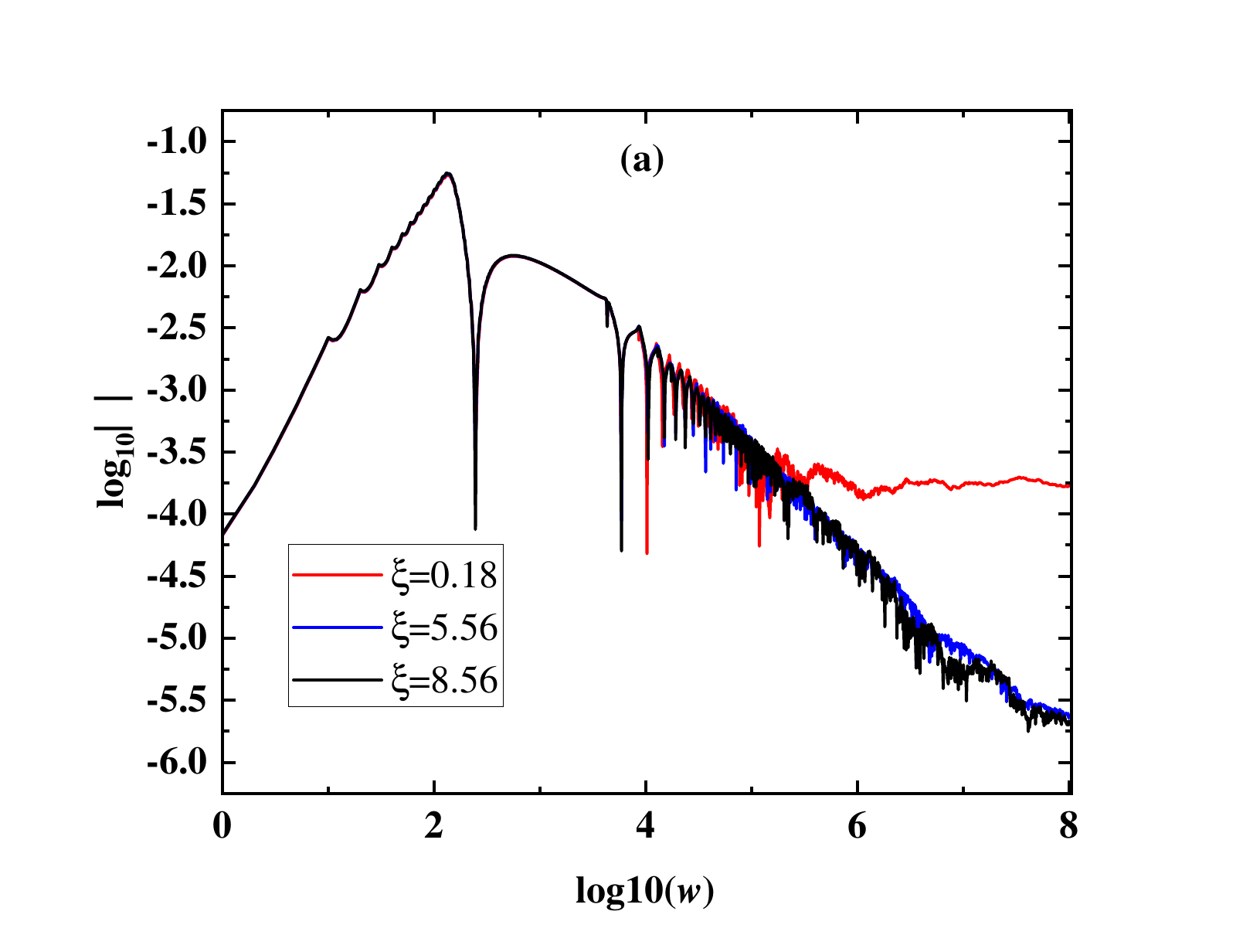}
        \includegraphics[width=12pc]{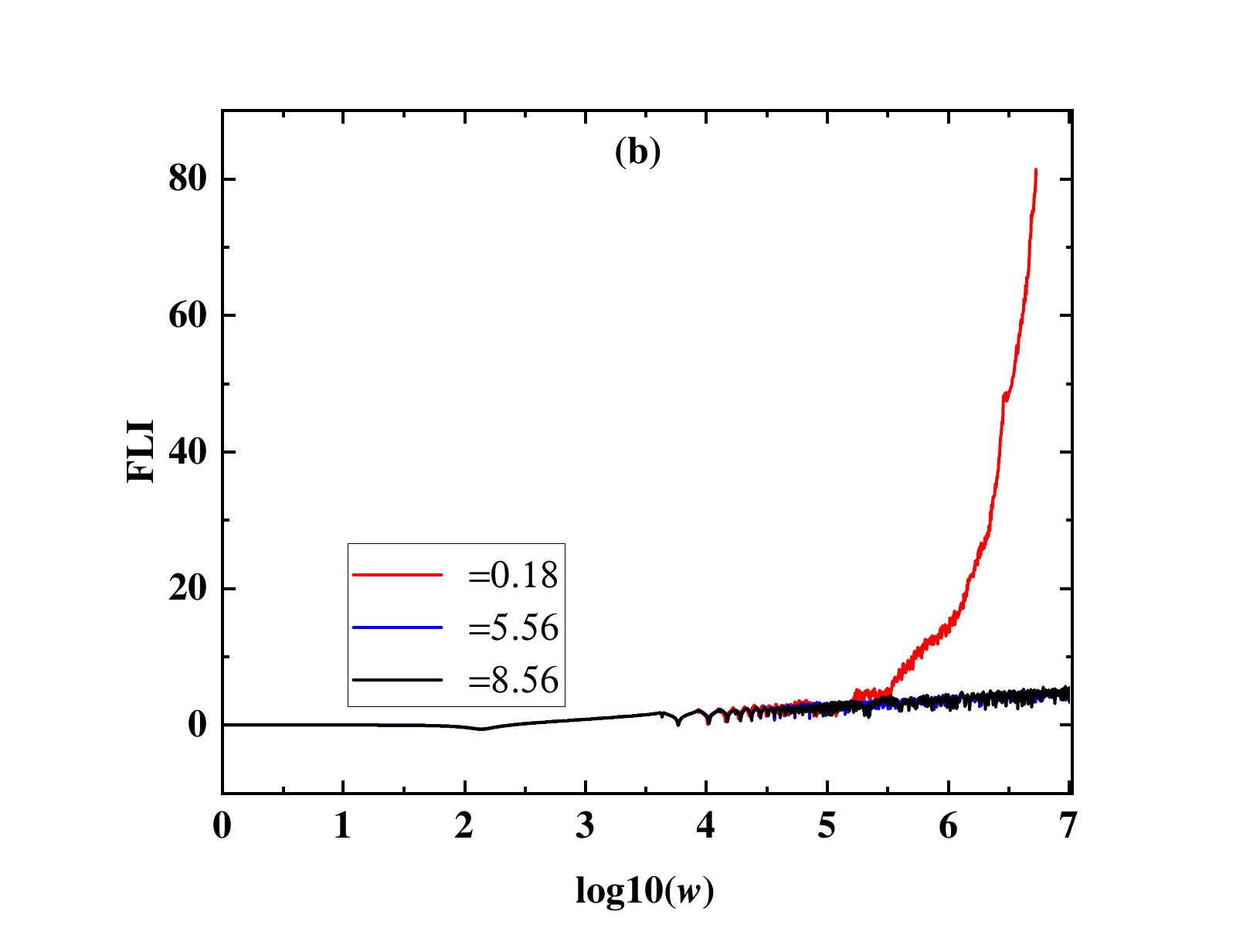}
        \includegraphics[width=12pc]{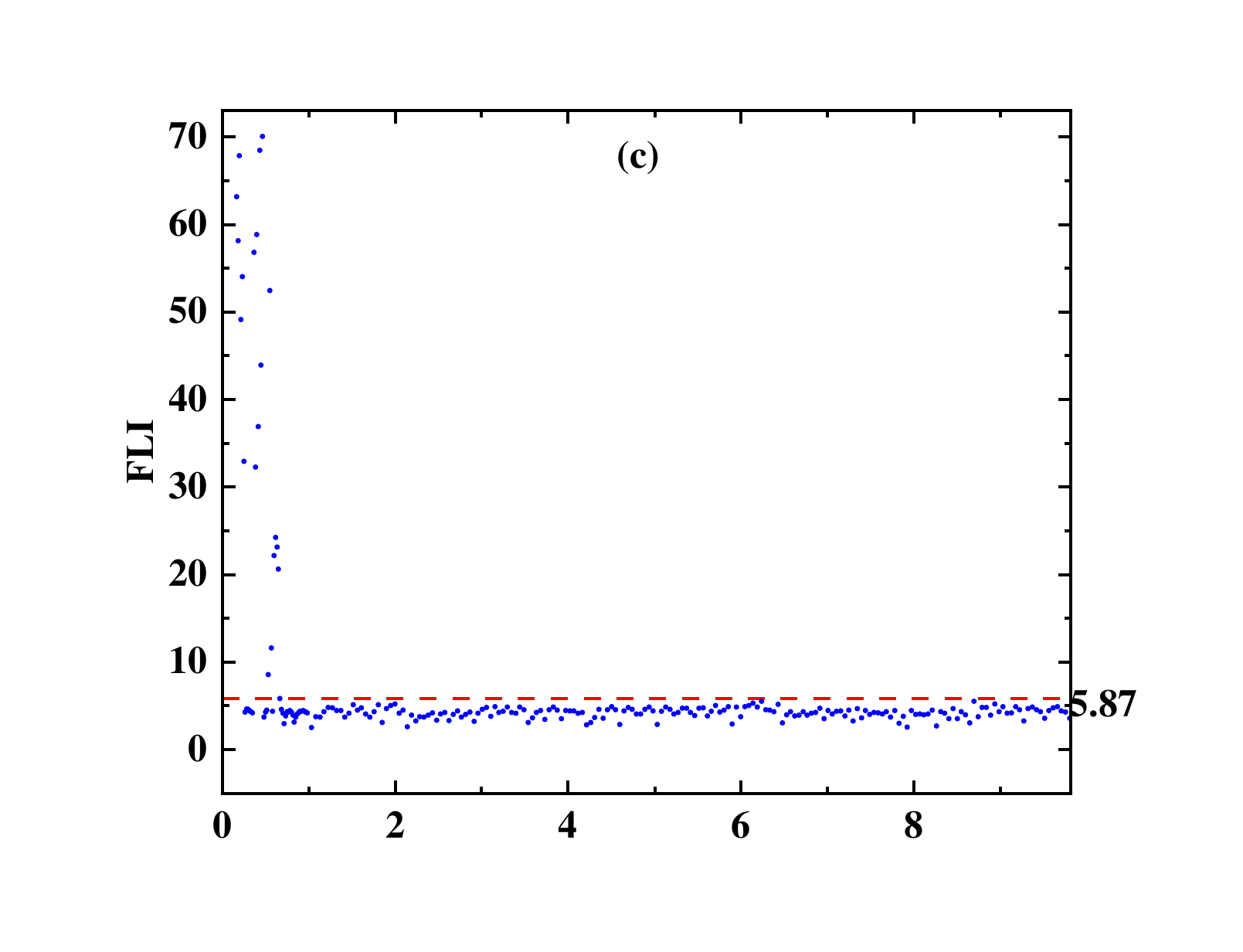}
\caption{(a) and (b): Same as Fig. 4 but both the maximal Lyapunov
exponents $\lambda$  and the FLIs are used. (c): Dependence of the
FLI on the parameter $\xi$ for the tested orbit in Fig. 4. For
$\xi=0.65$, FLI=5.87 is a threshold between order and chaos.
        }
    }
\end{figure*}

\begin{figure*}[htpb]
    \centering{
        \includegraphics[width=17pc]{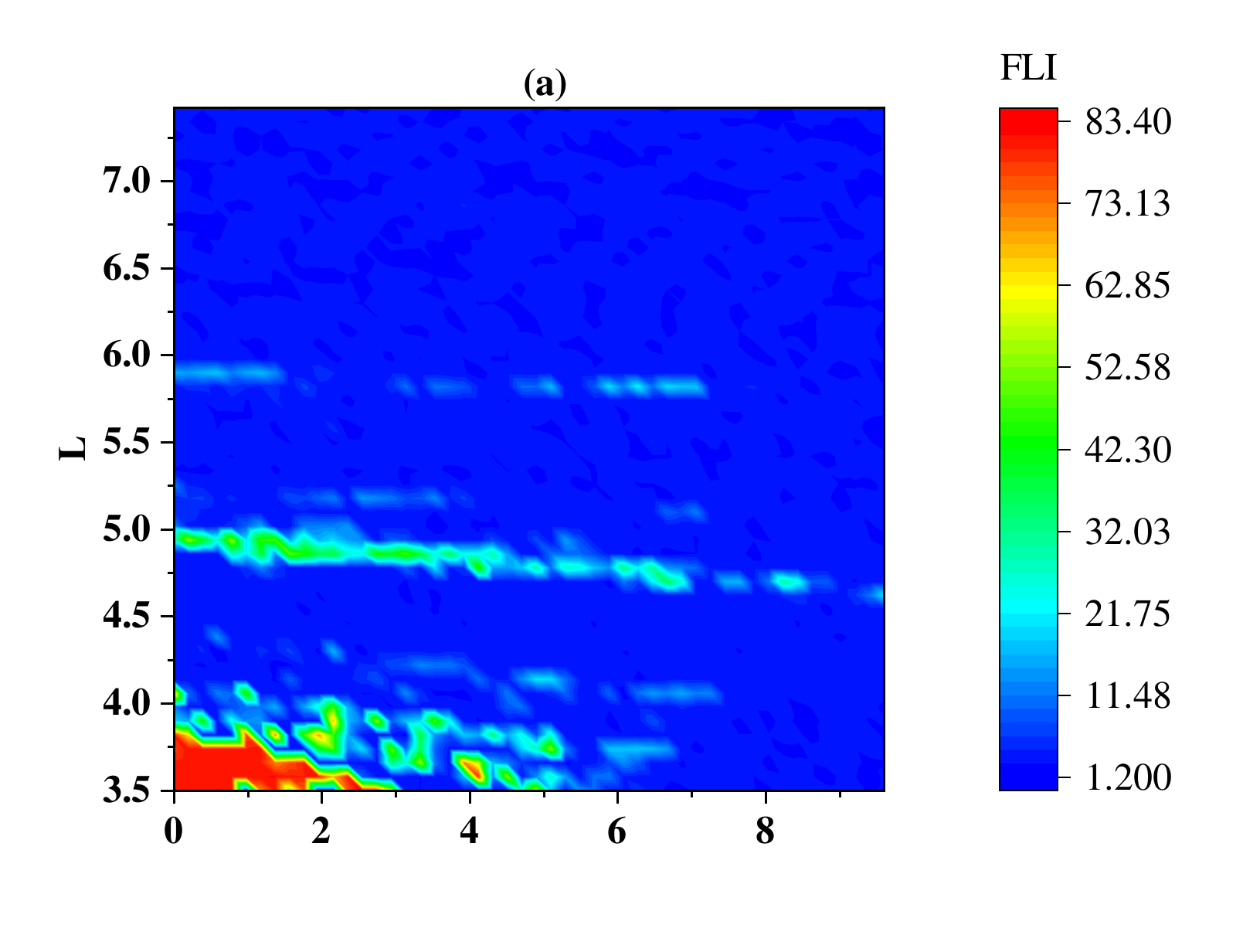}
        \includegraphics[width=17pc]{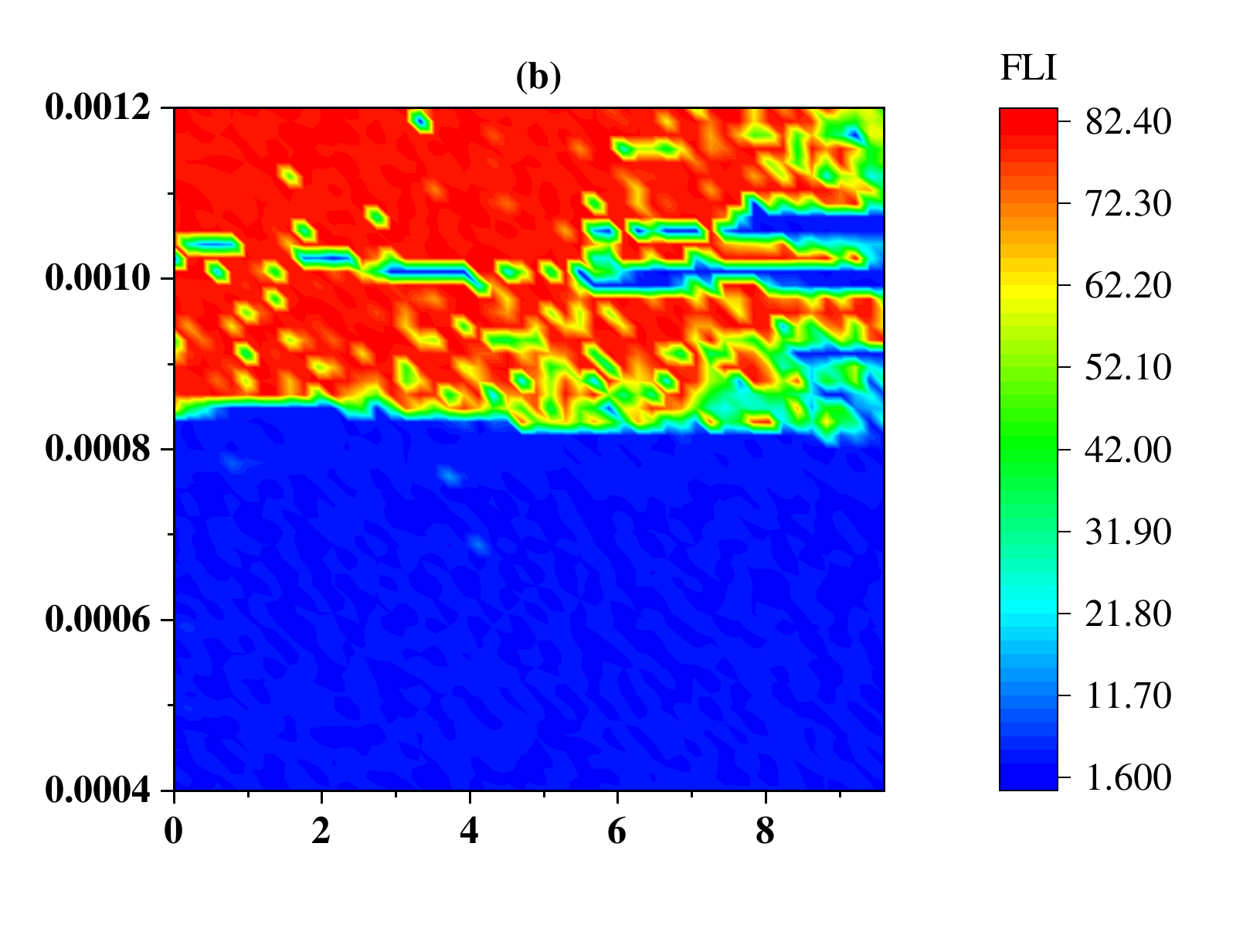}
        \includegraphics[width=17pc]{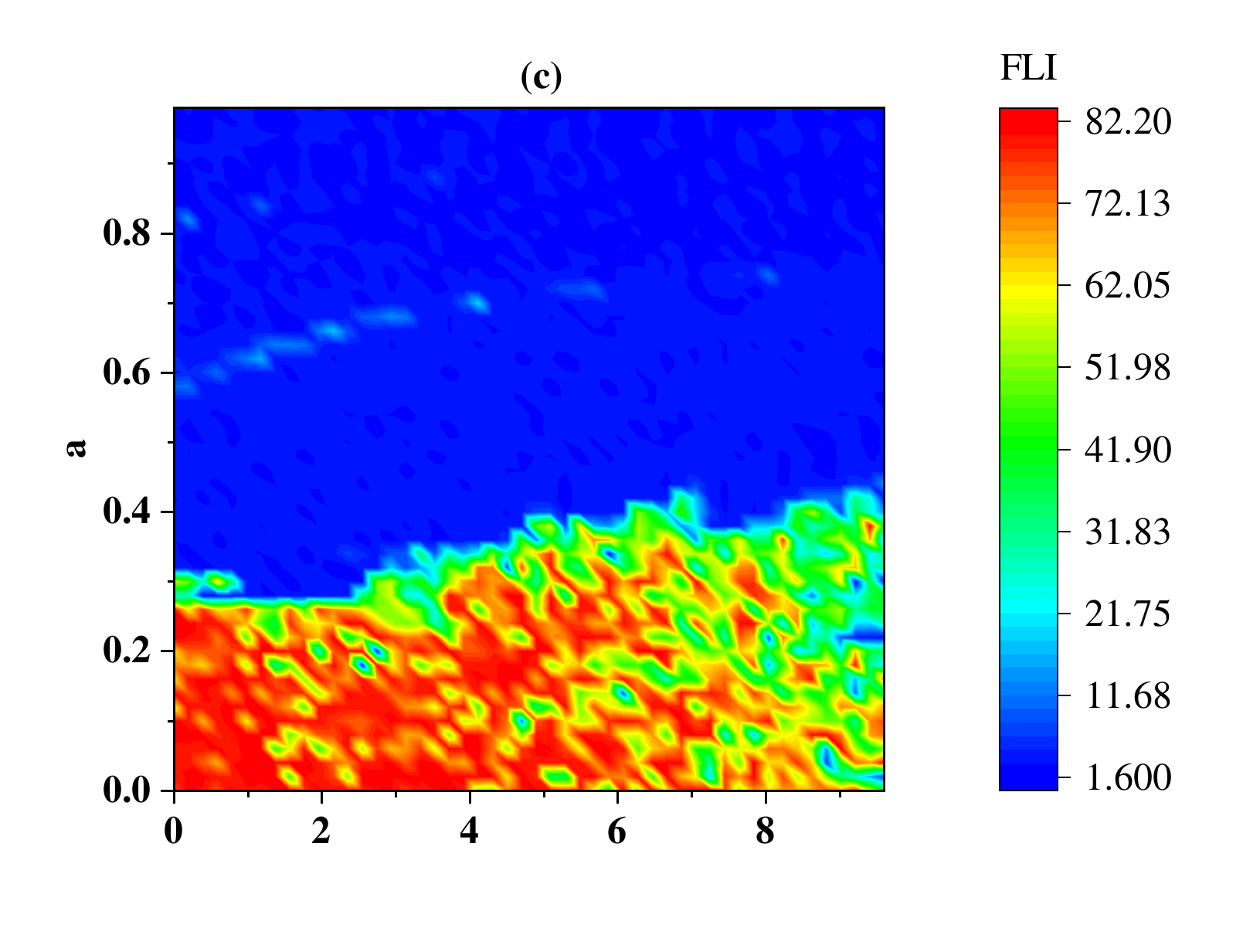}
        \includegraphics[width=17pc]{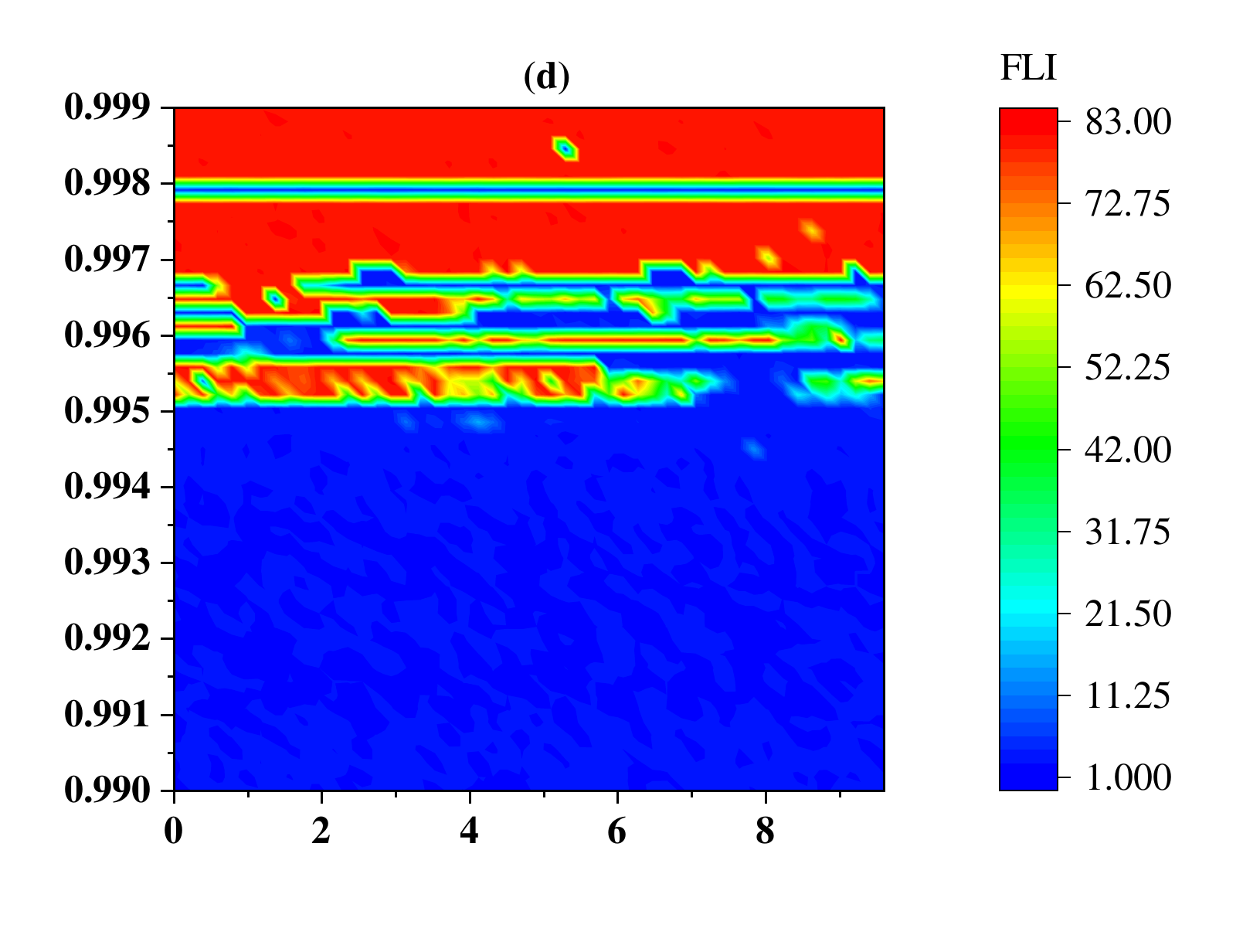}
\caption{Distributions of two parameters for order and chaos,
which are described by the FLIs. The tested orbit has its  initial
radius $r=70$. (a) The two-parameter space $(\xi,L)$, where the
other parameters are $E=0.995$, $a=0.5$ and $\beta=8\times
10^{-4}$. (b) The two-parameter space $(\xi,\beta)$, where the
other parameters are $E=0.995$, $L=4.5$ and $a=0.5$. (c) The
two-parameter space $(\xi,a)$, where the other parameters are
$E=0.995$, $L=4.5$ and $\beta=8\times 10^{-4}$. (d) The
two-parameter space $(\xi,E)$, where the other parameters are
$a=0.5$, $L=4.5$ and $\beta=8\times 10^{-4}$.
        }
    }
\end{figure*}

\begin{figure*}[htpb]
    \centering{
        \includegraphics[width=17pc]{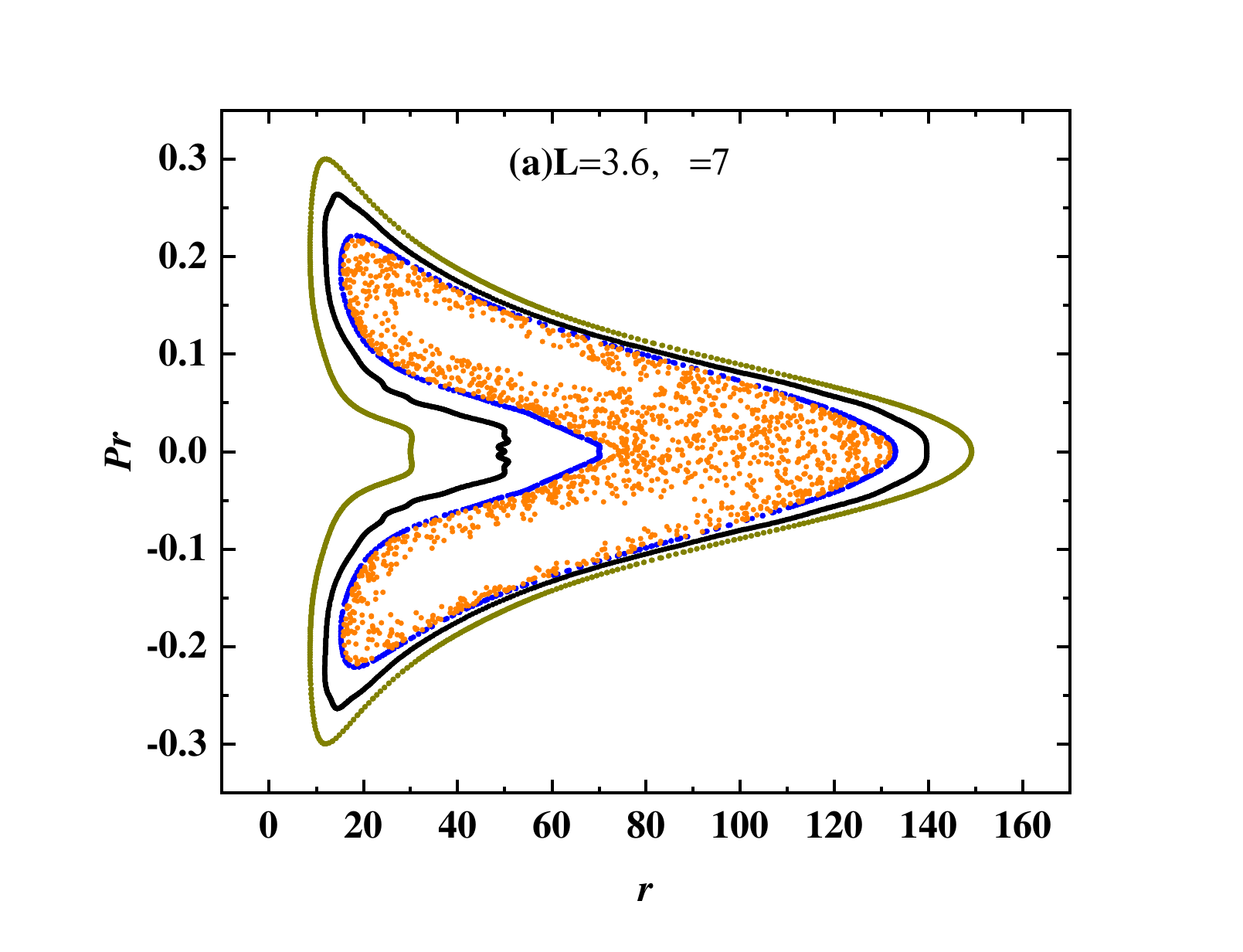}
        \includegraphics[width=17pc]{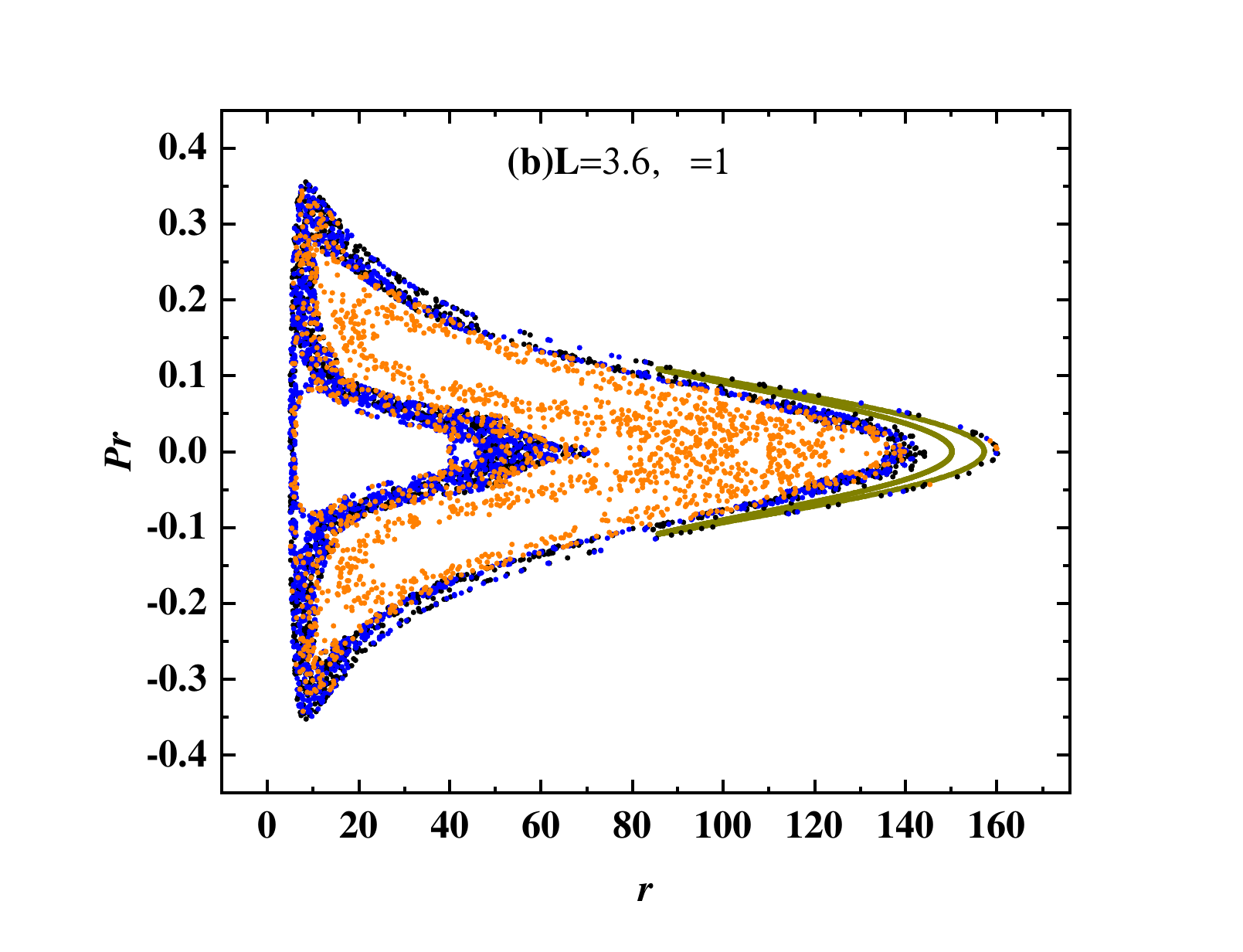}
        \includegraphics[width=17pc]{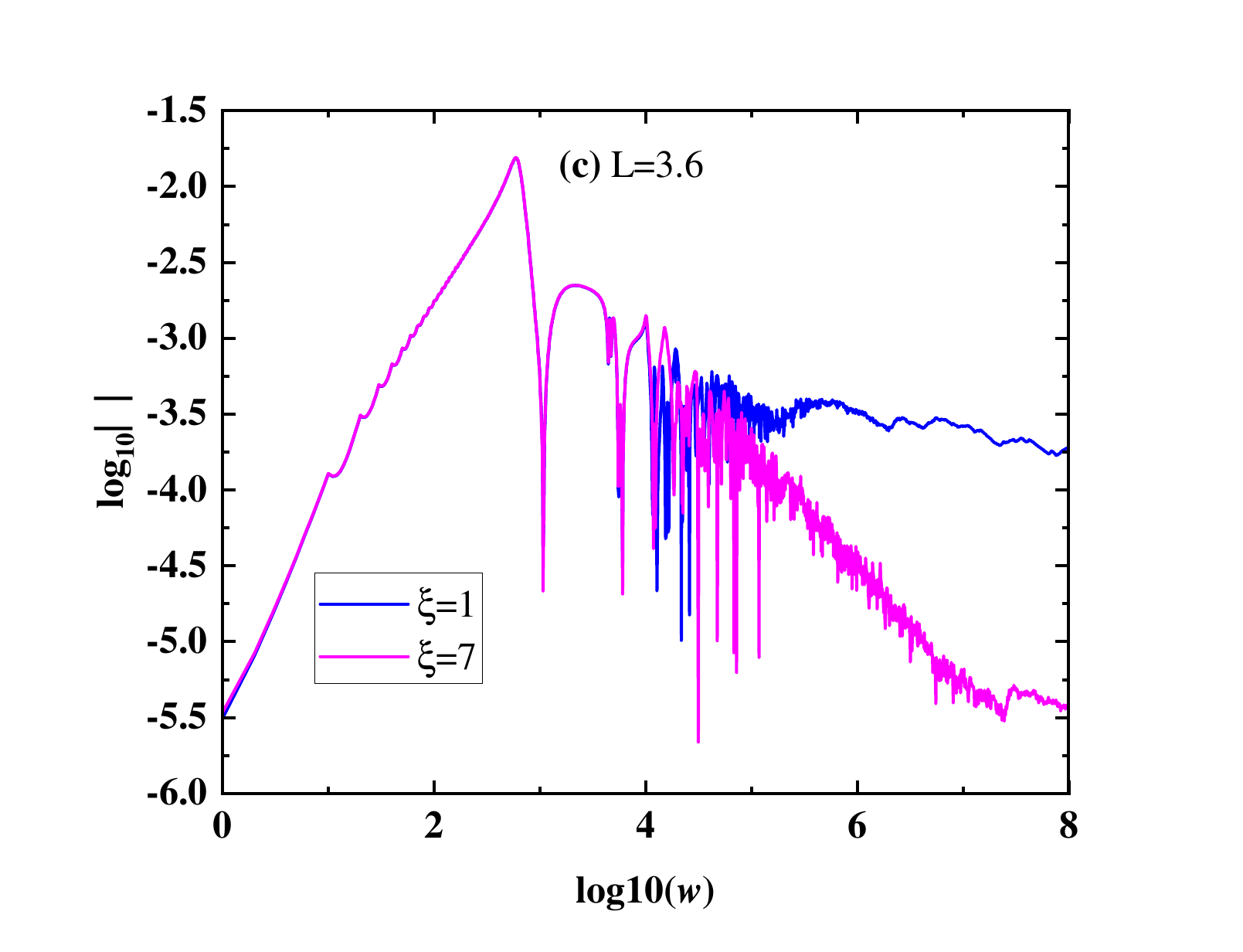}
        \includegraphics[width=17pc]{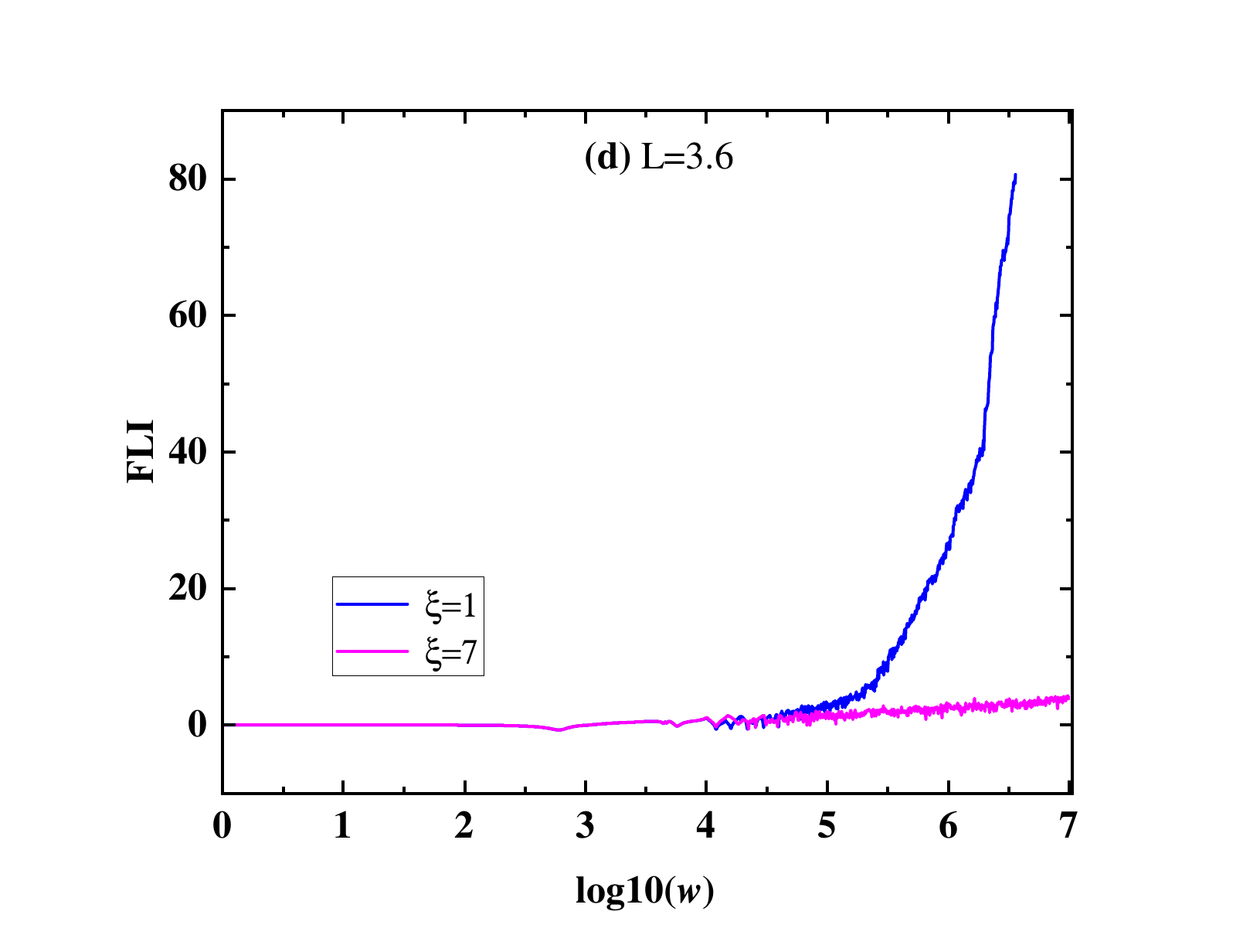}
        \includegraphics[width=17pc]{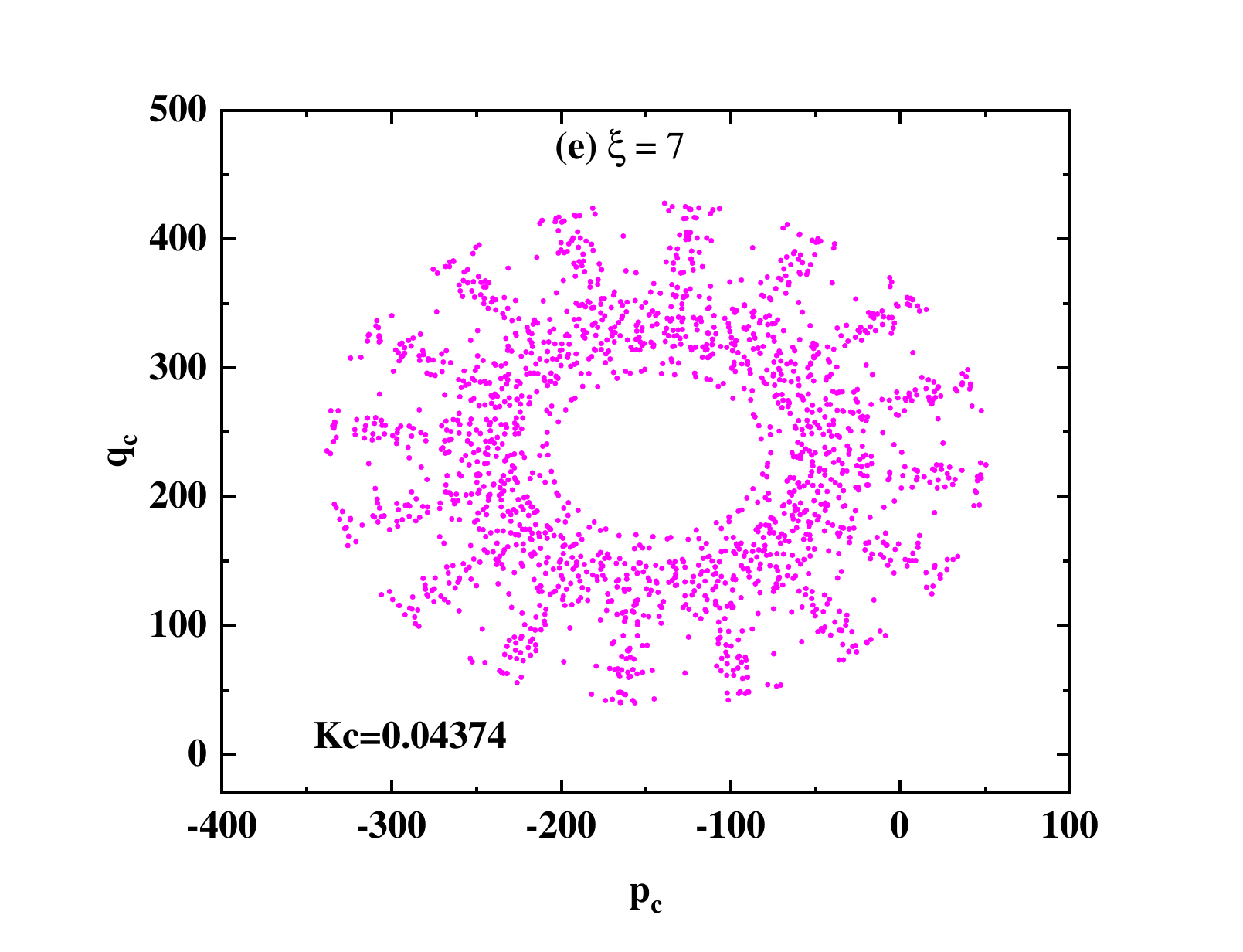}
        \includegraphics[width=17pc]{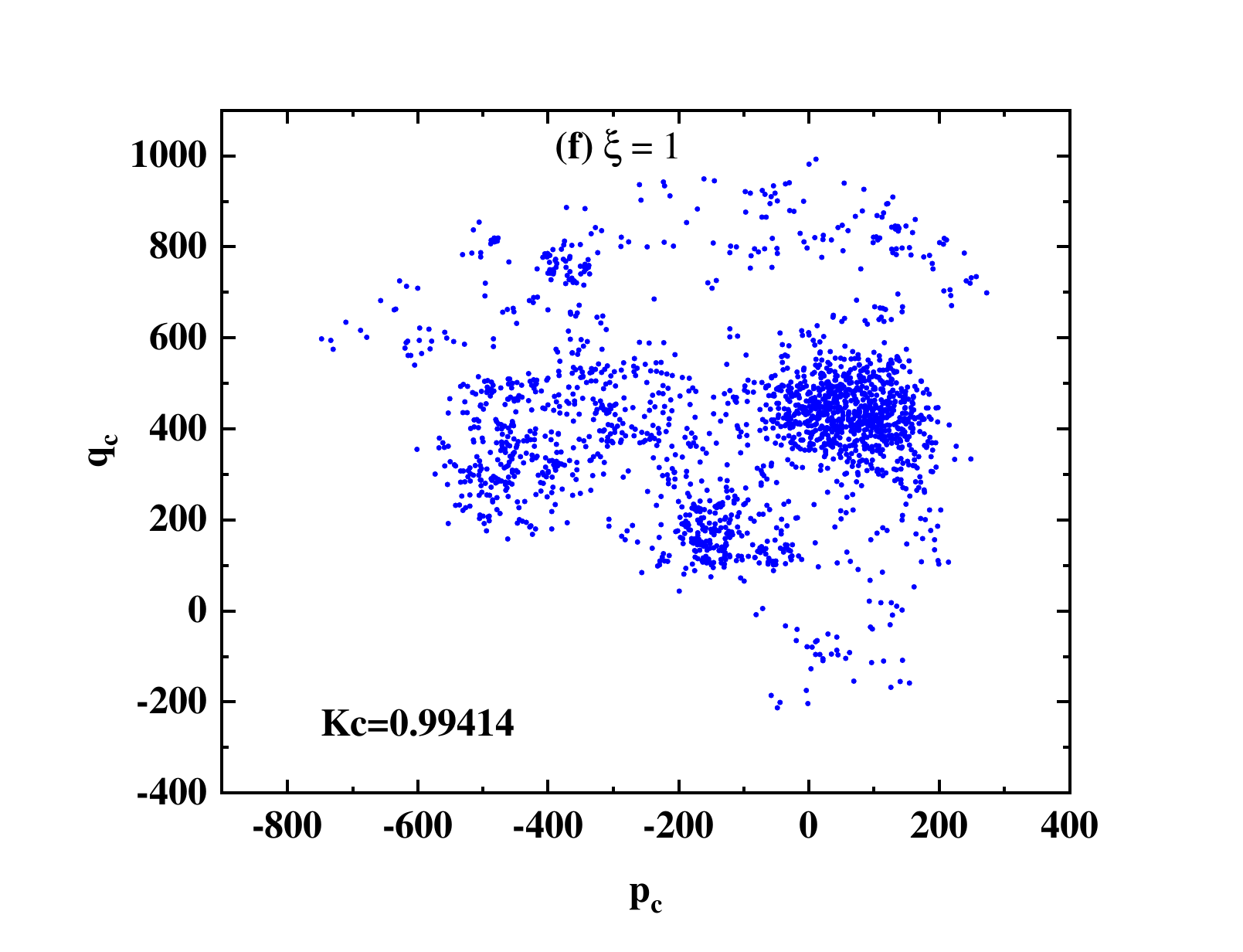}
\caption{Dynamical behaviors for two values of the parameter $\xi$
in Fig. 6(a). (a): Poincar\'{e} sections for $\xi=7$. (b):
Poincar\'{e} sections for $\xi=1$. (c): The maximal Lyapunov
exponents $\lambda$ of the tested orbit with the initial
separation $r=70$ for the  two values of $\xi$. (d): The FLIs of
the tested orbit for the  two values of $\xi$. (e): The visual of
$(p_c, q_c)$ and the value of $K_{c}$ of the tested orbit for
$\xi=7$. (f): The visual of $(p_c, q_c)$ and the value of $K_{c}$
of the tested orbit for $\xi=1$.
        }
    }
\end{figure*}

\begin{figure*}[htpb]
    \centering{
        \includegraphics[width=17pc]{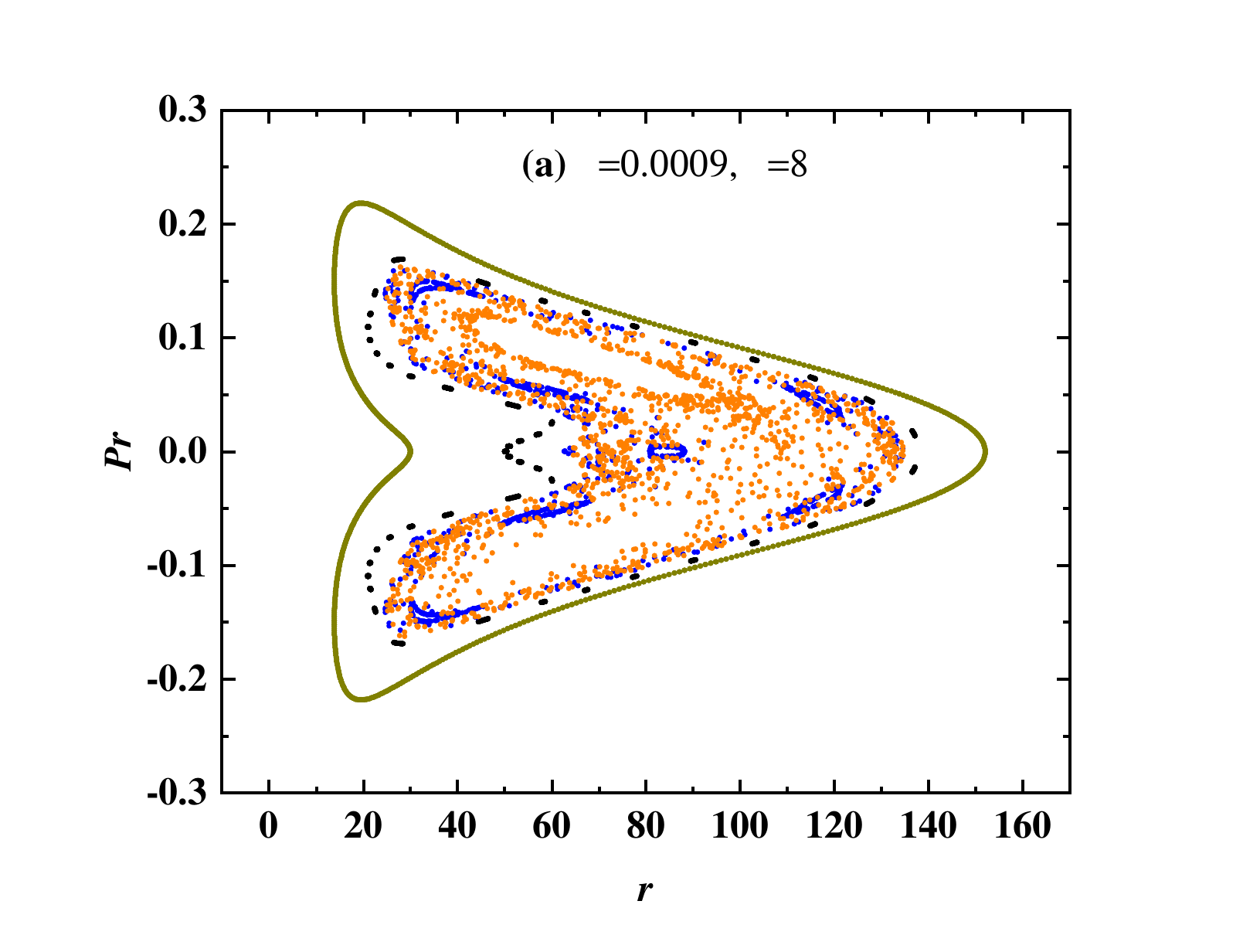}
        \includegraphics[width=17pc]{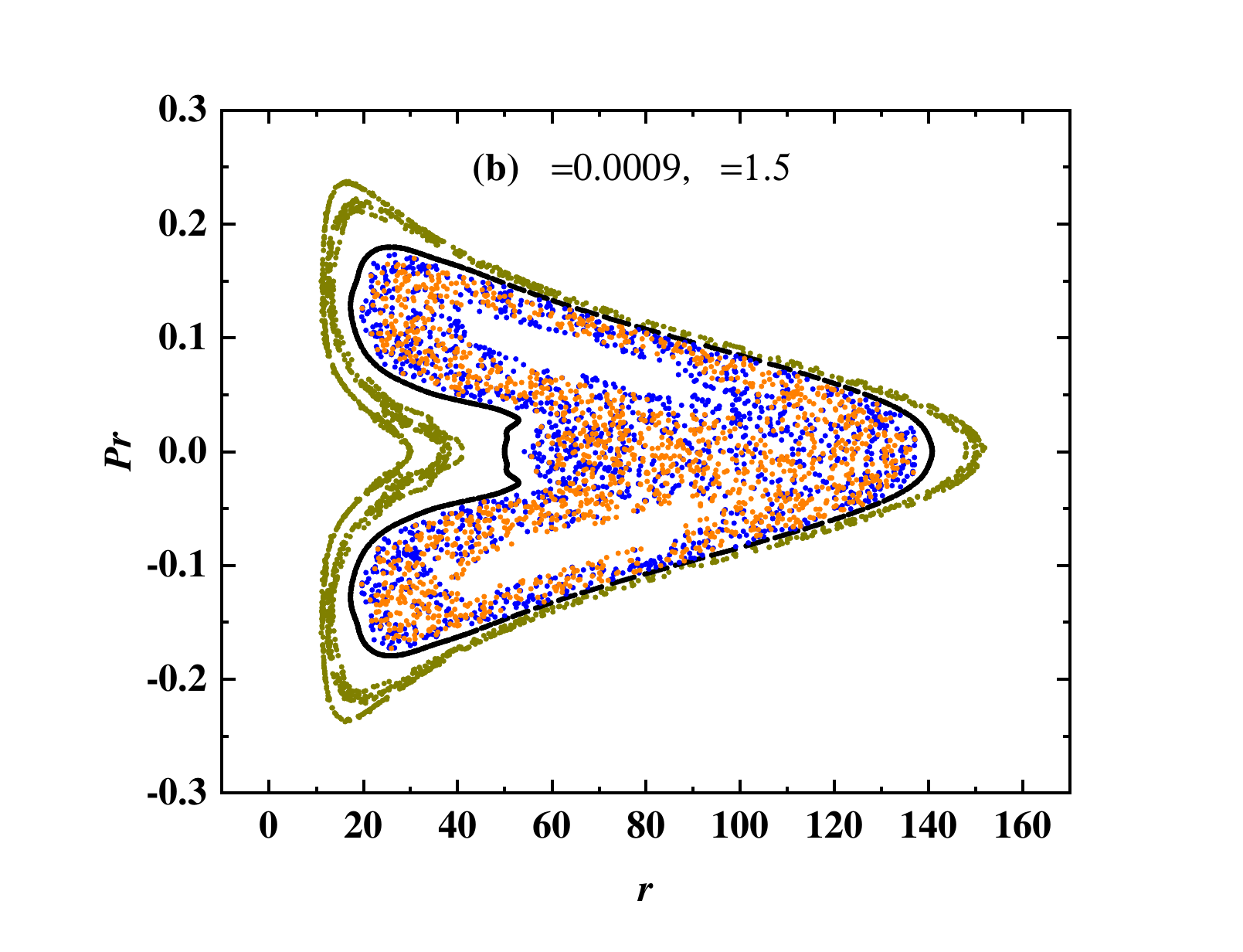}
        \includegraphics[width=17pc]{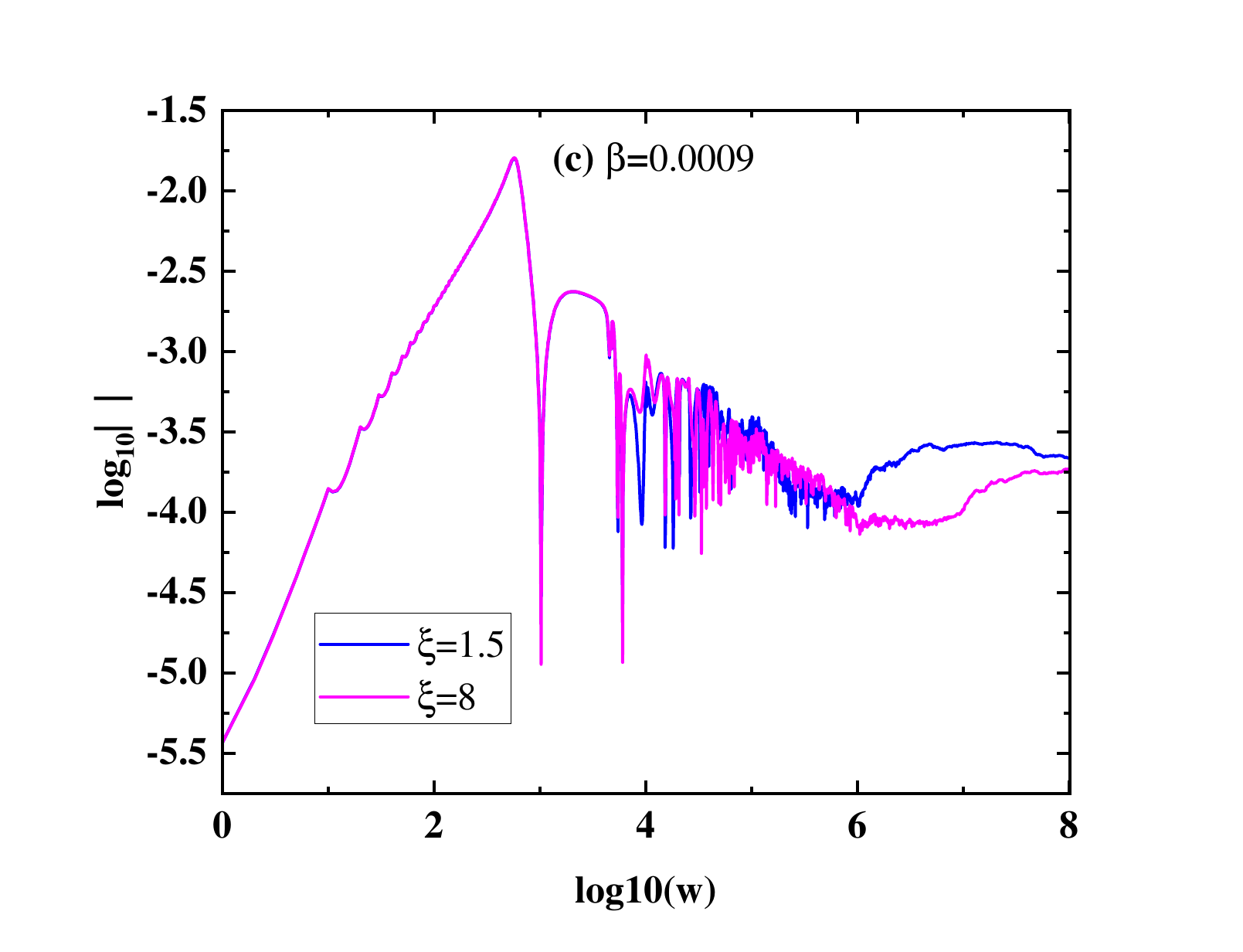}
        \includegraphics[width=17pc]{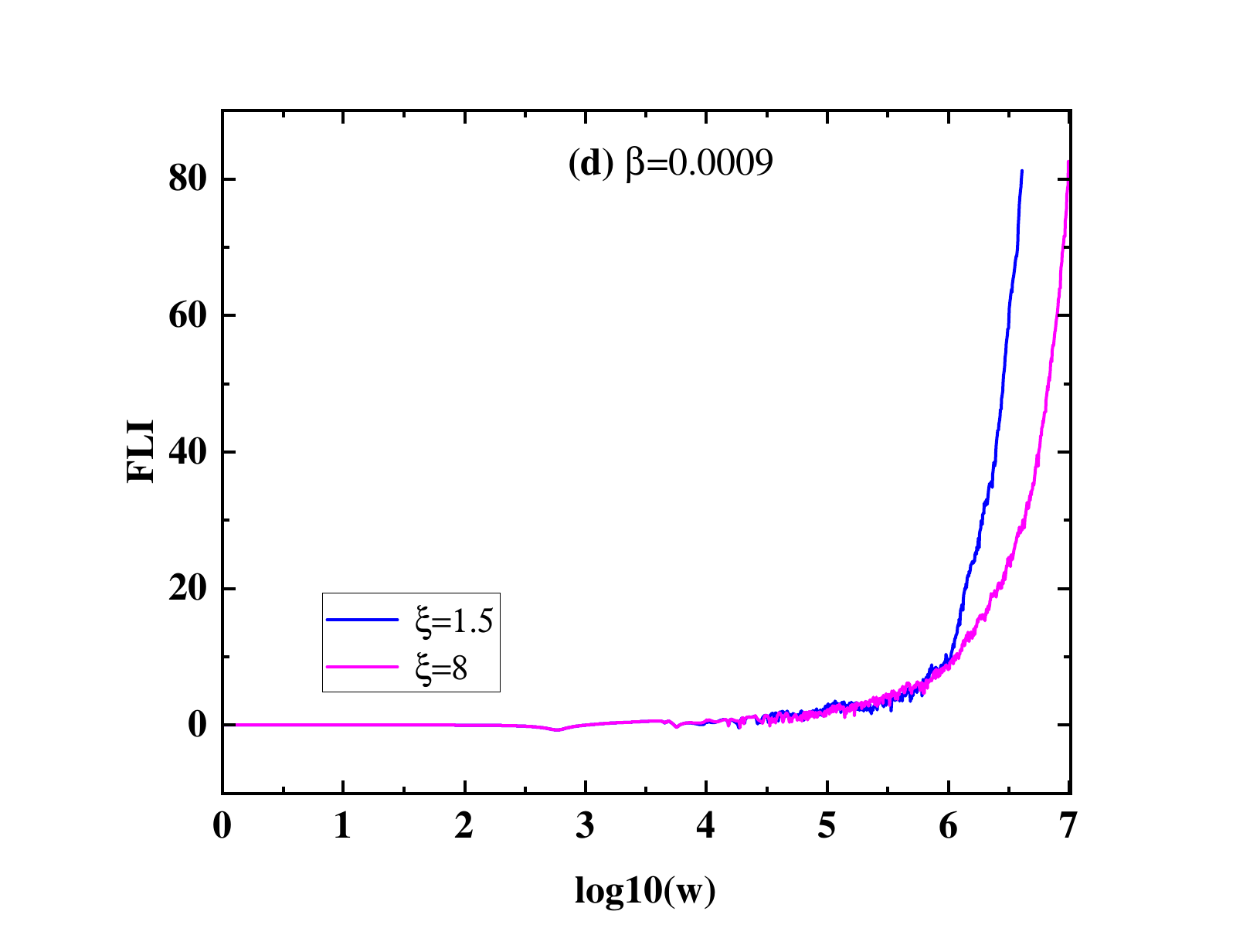}
        \includegraphics[width=17pc]{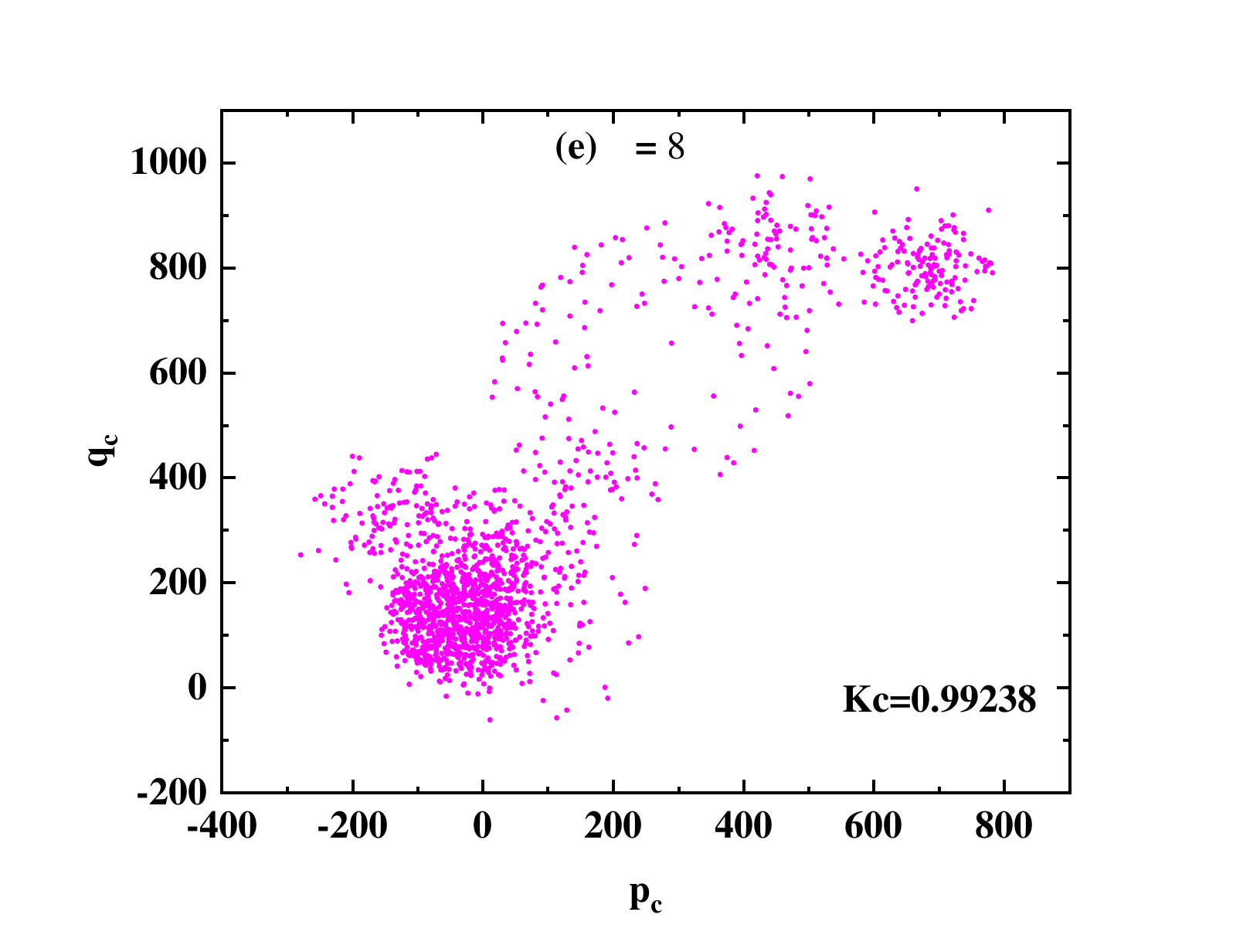}
        \includegraphics[width=17pc]{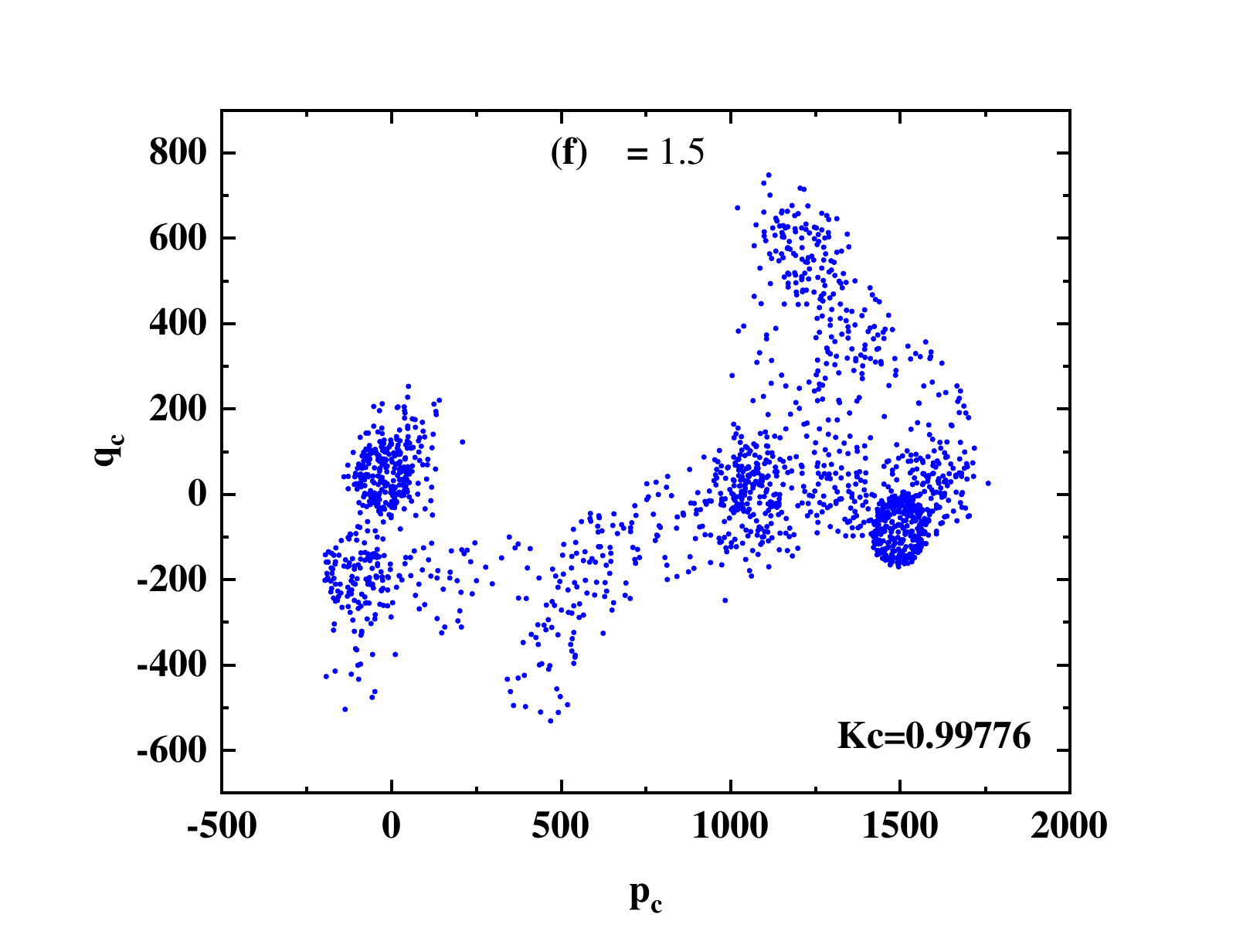}
\caption{Same as Fig. 7 but two values of the parameter $\xi$ in
Fig. 6(b) are considered.
        }
    }
\end{figure*}

\begin{figure*}[htpb]
    \centering{
        \includegraphics[width=17pc]{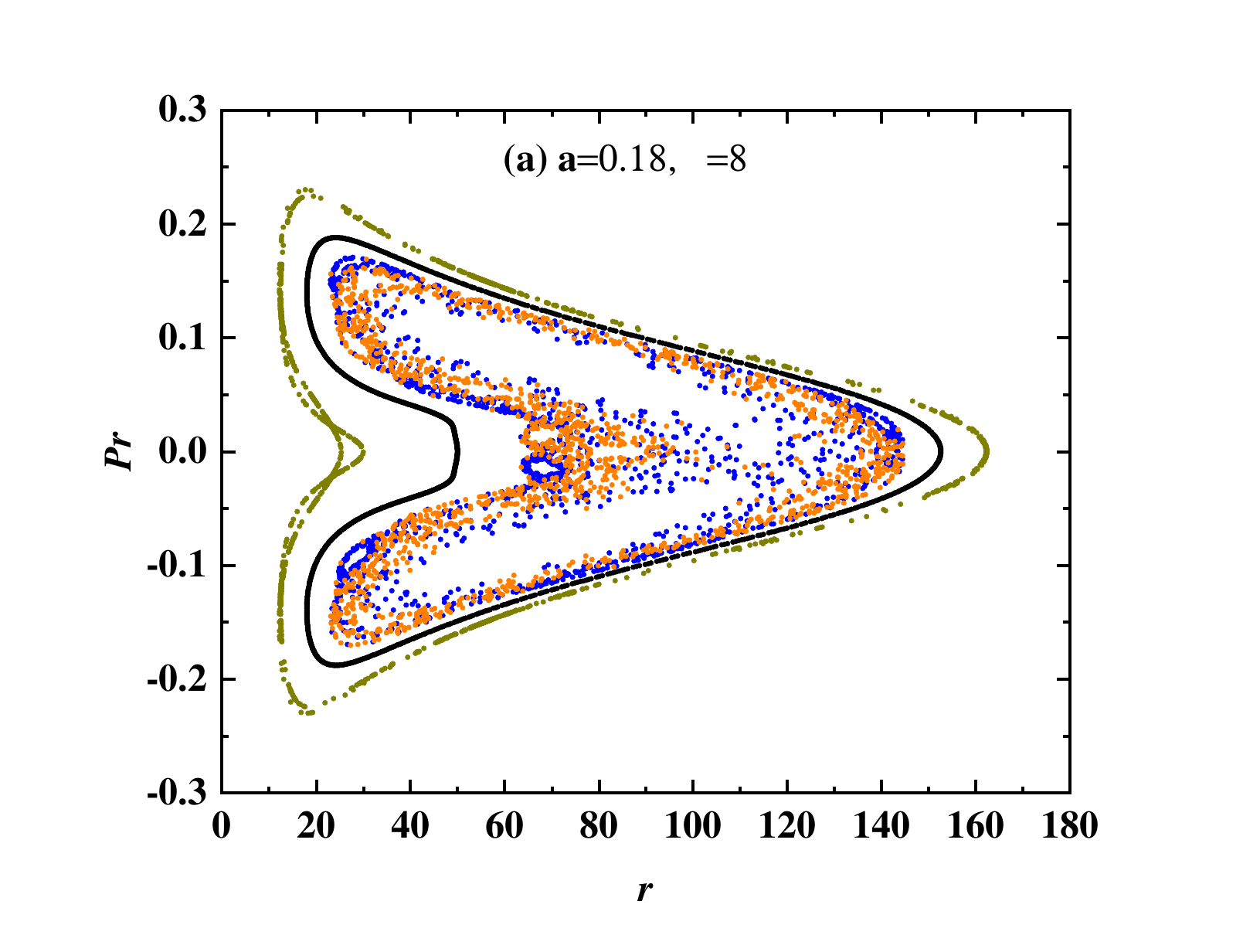}
        \includegraphics[width=17pc]{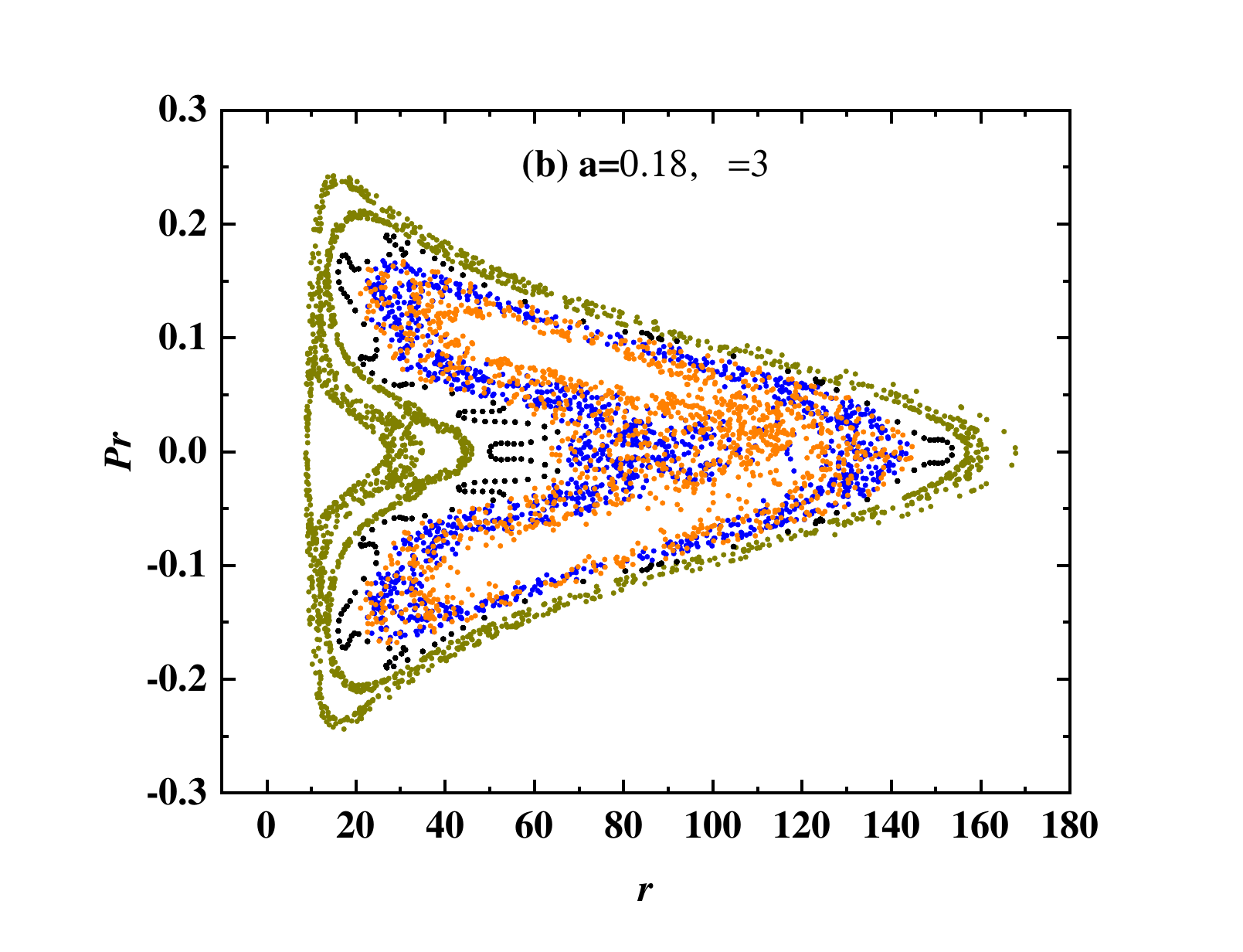}
        \includegraphics[width=17pc]{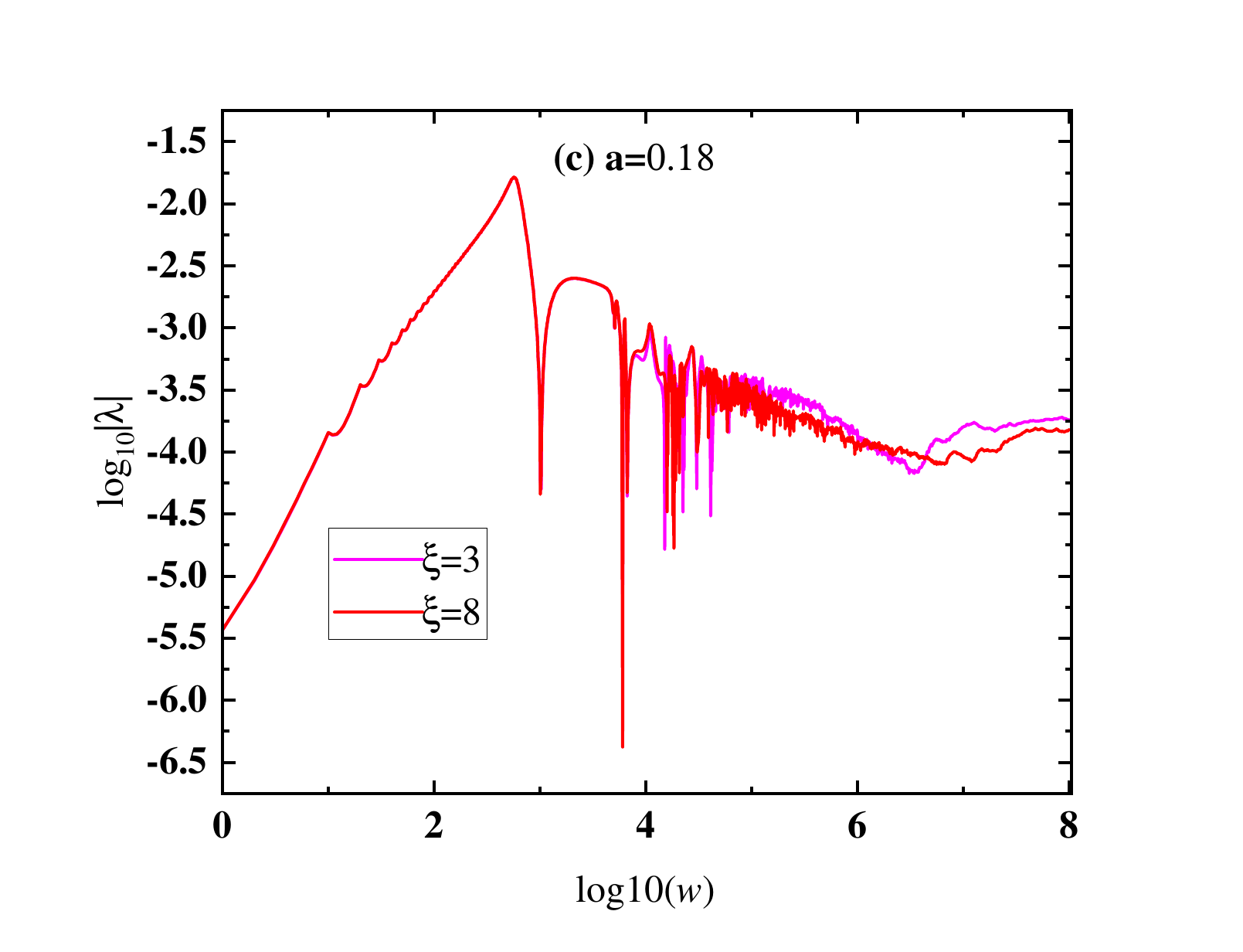}
        \includegraphics[width=17pc]{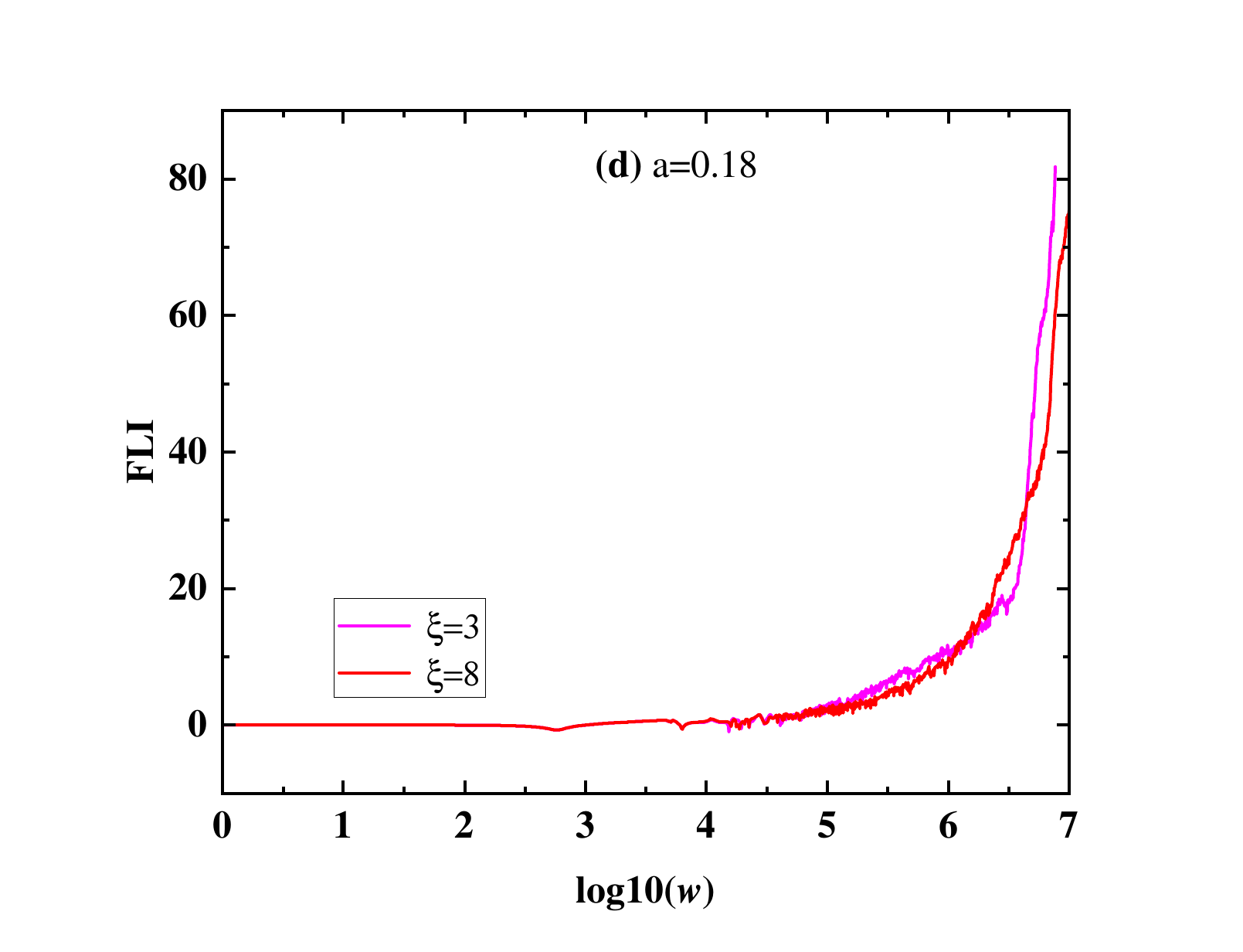}
        \includegraphics[width=17pc]{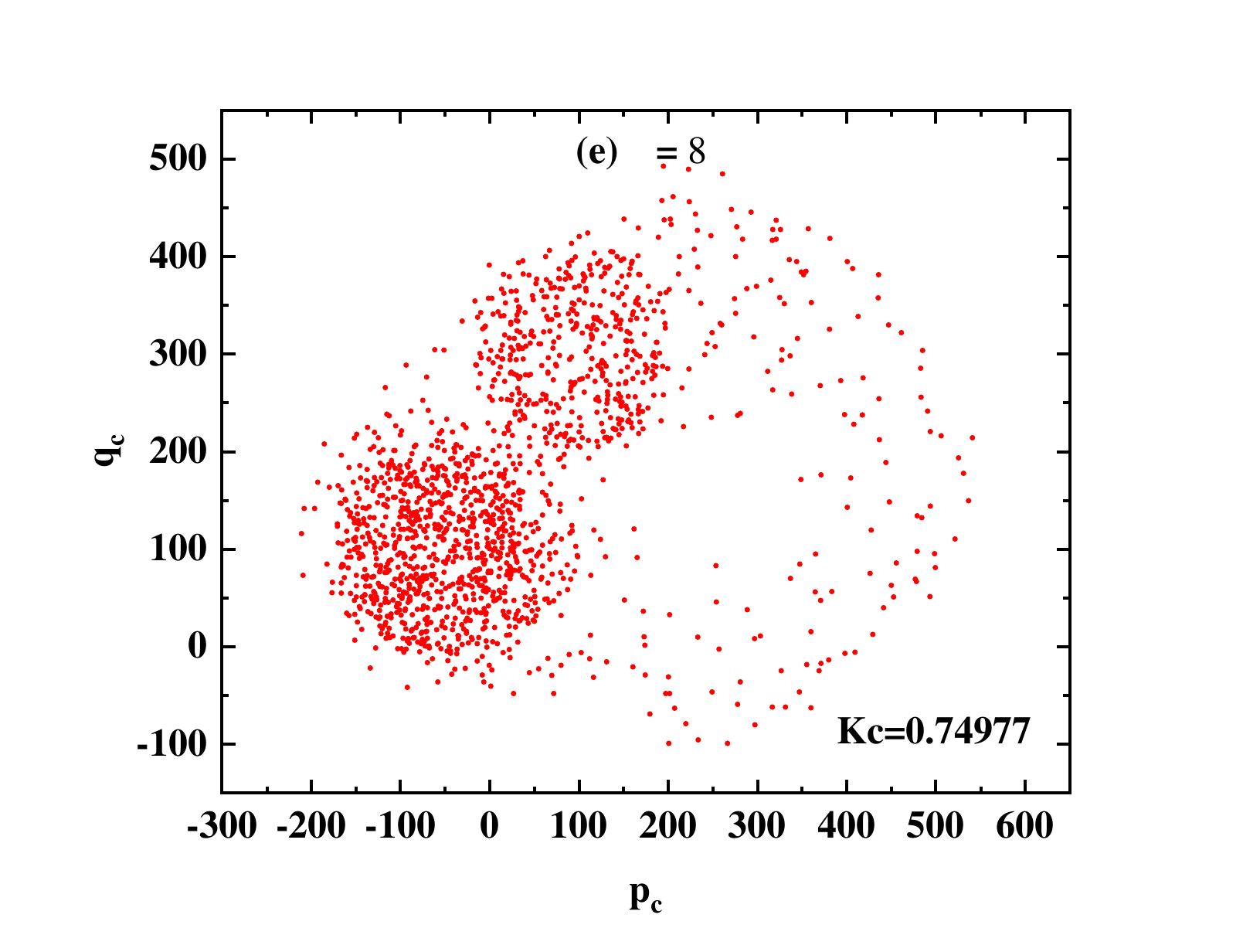}
        \includegraphics[width=17pc]{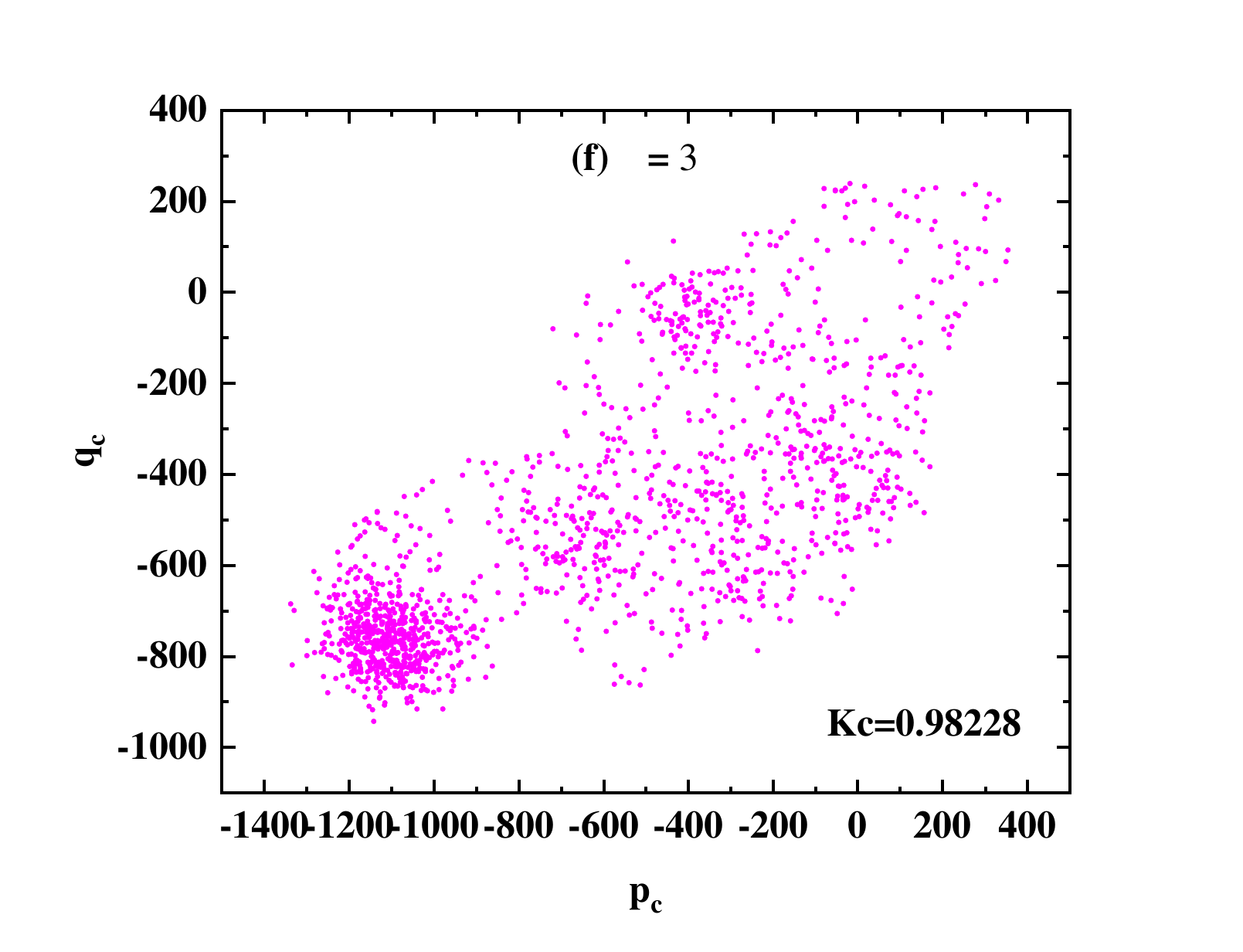}
\caption{Same as Fig. 7 but two values of the parameter $\xi$ in
Fig. 6(c) are chosen.
        }
    }
\end{figure*}

\begin{figure*}[htpb]
    \centering{
        \includegraphics[width=17pc]{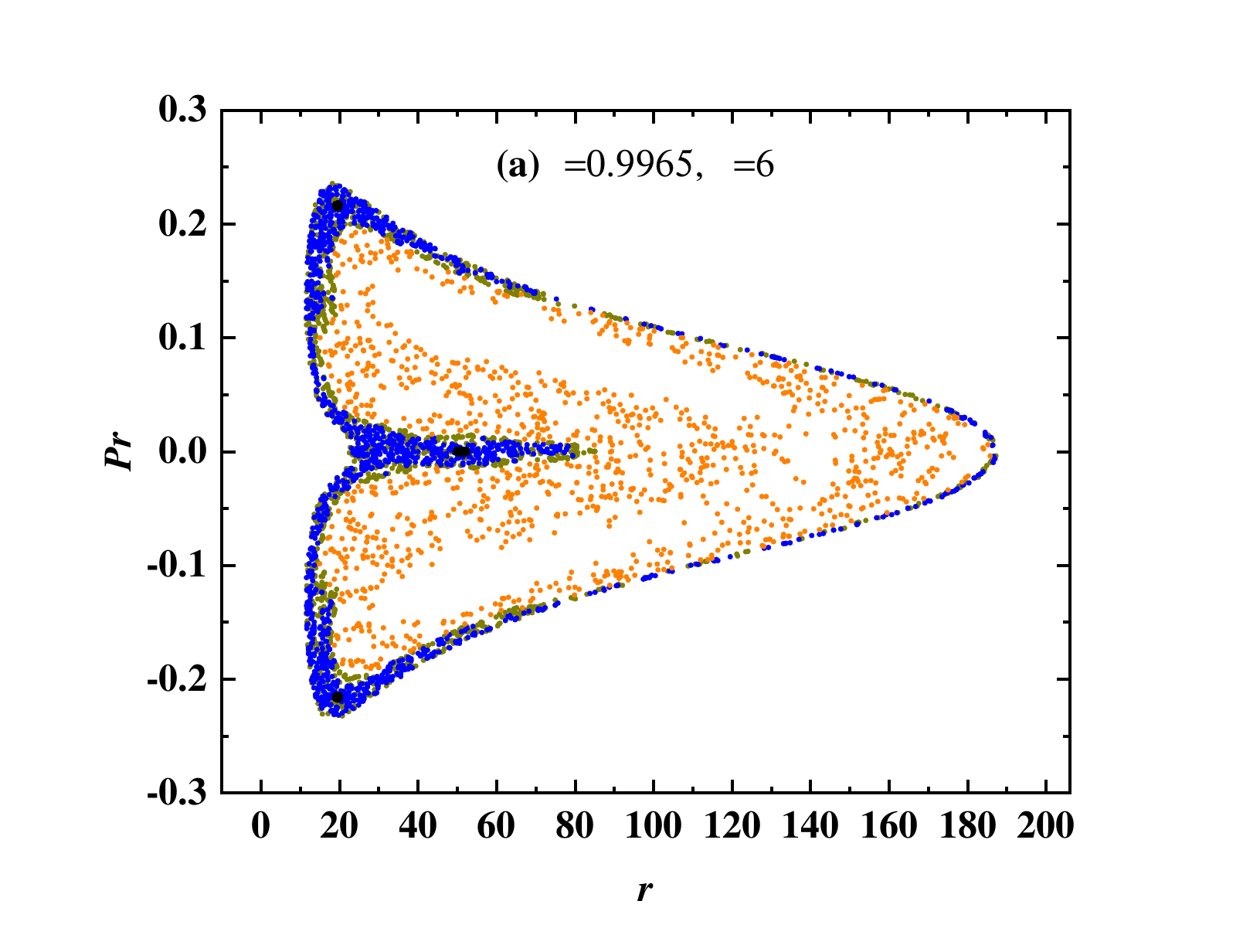}
        \includegraphics[width=17pc]{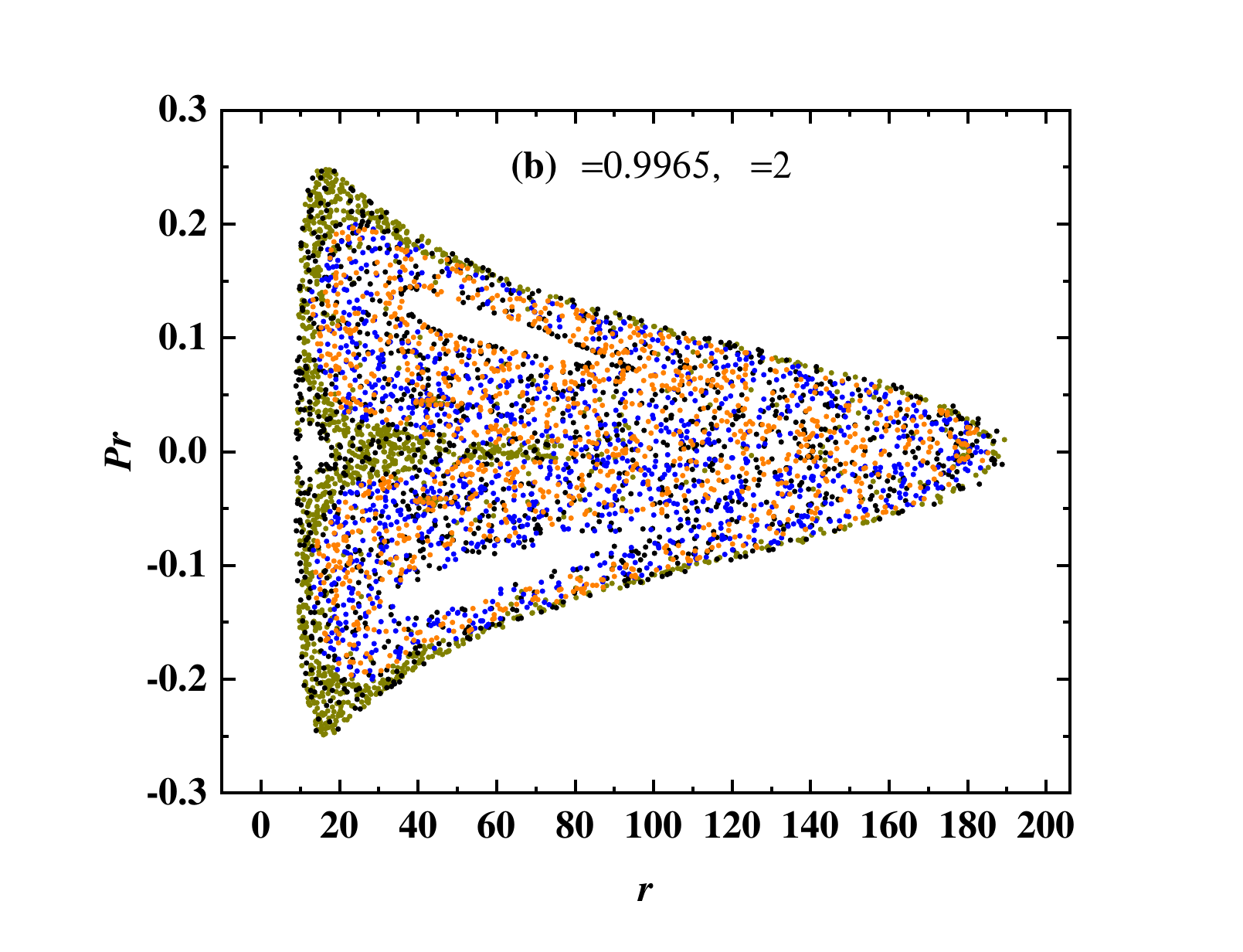}
        \includegraphics[width=17pc]{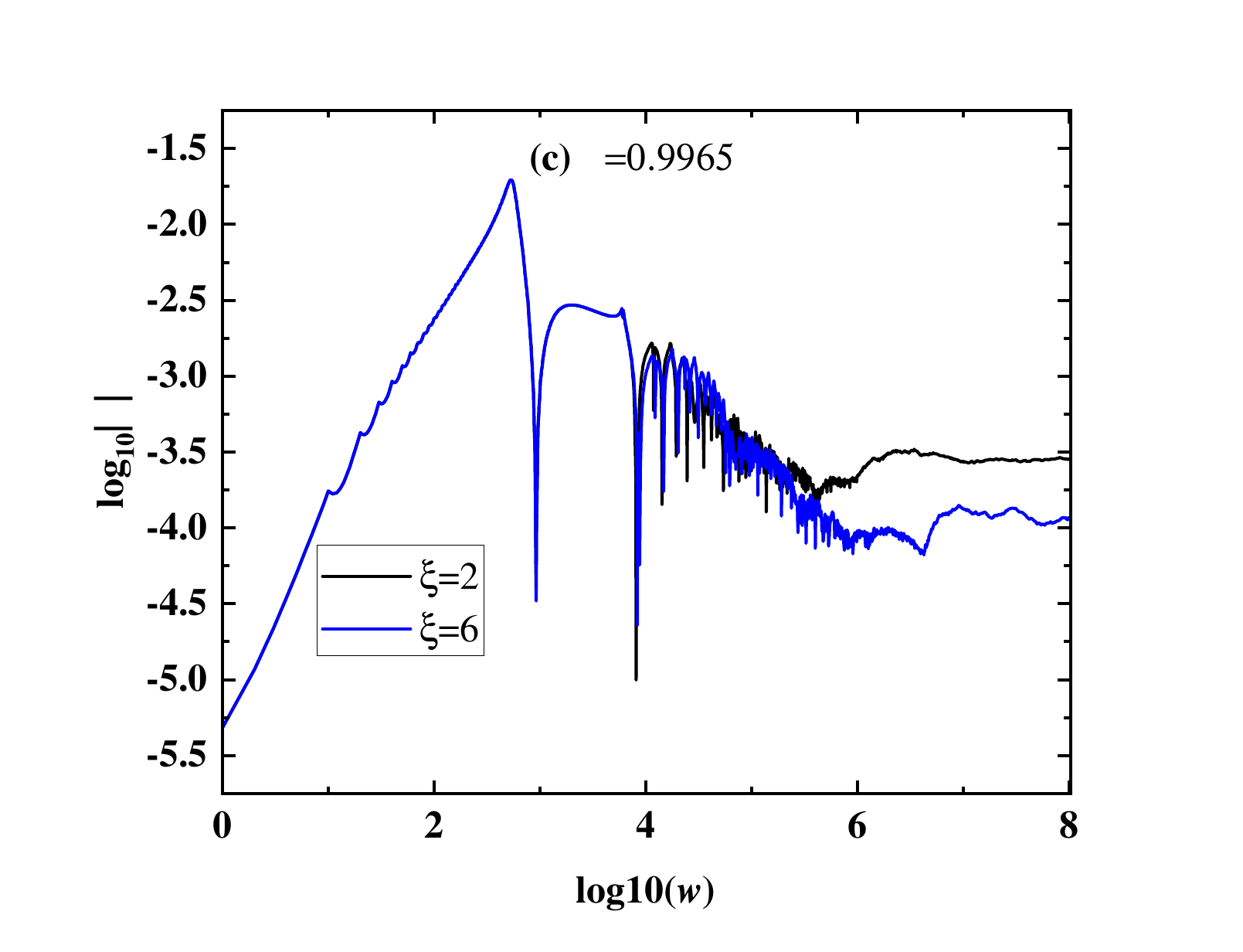}
        \includegraphics[width=17pc]{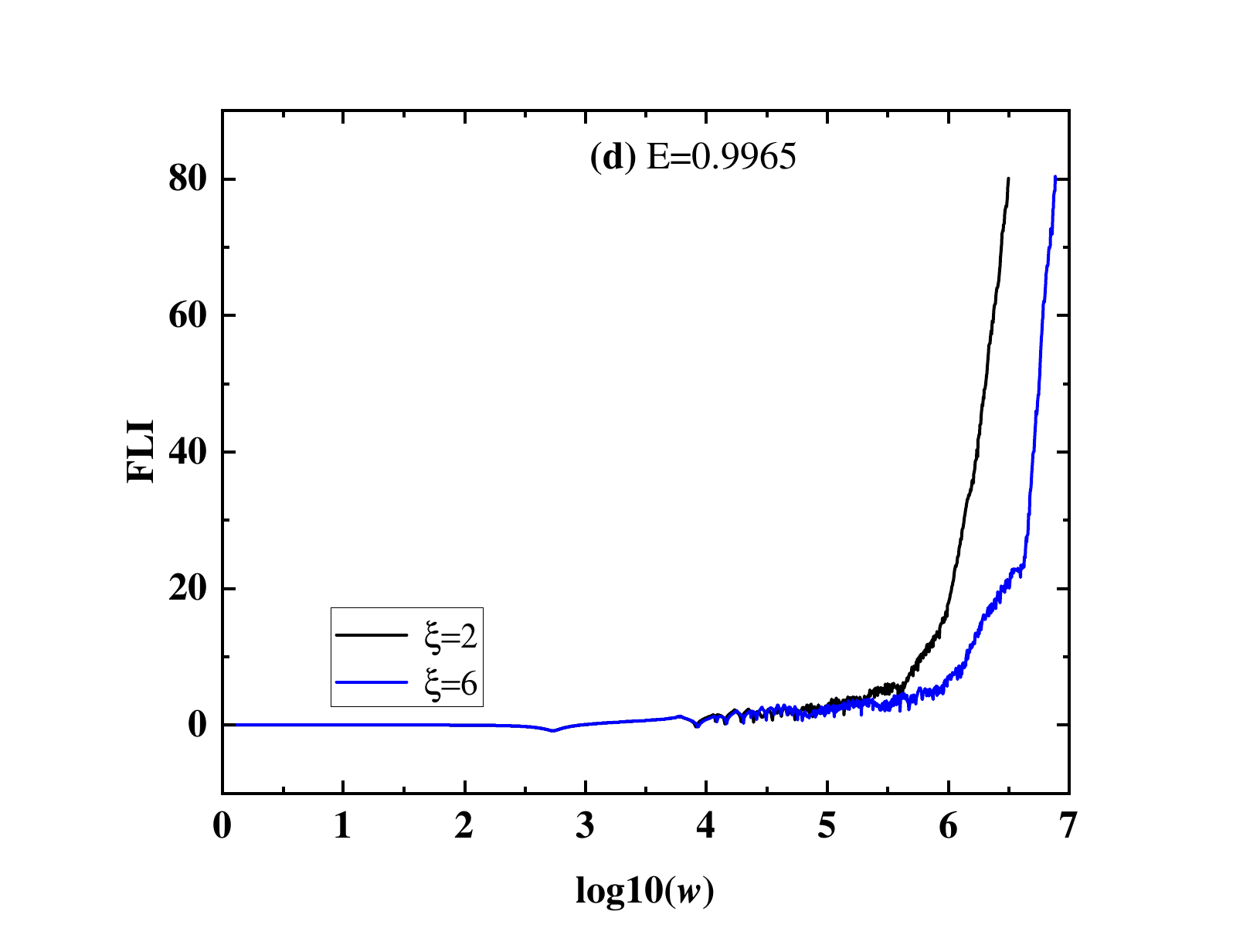}
        \includegraphics[width=17pc]{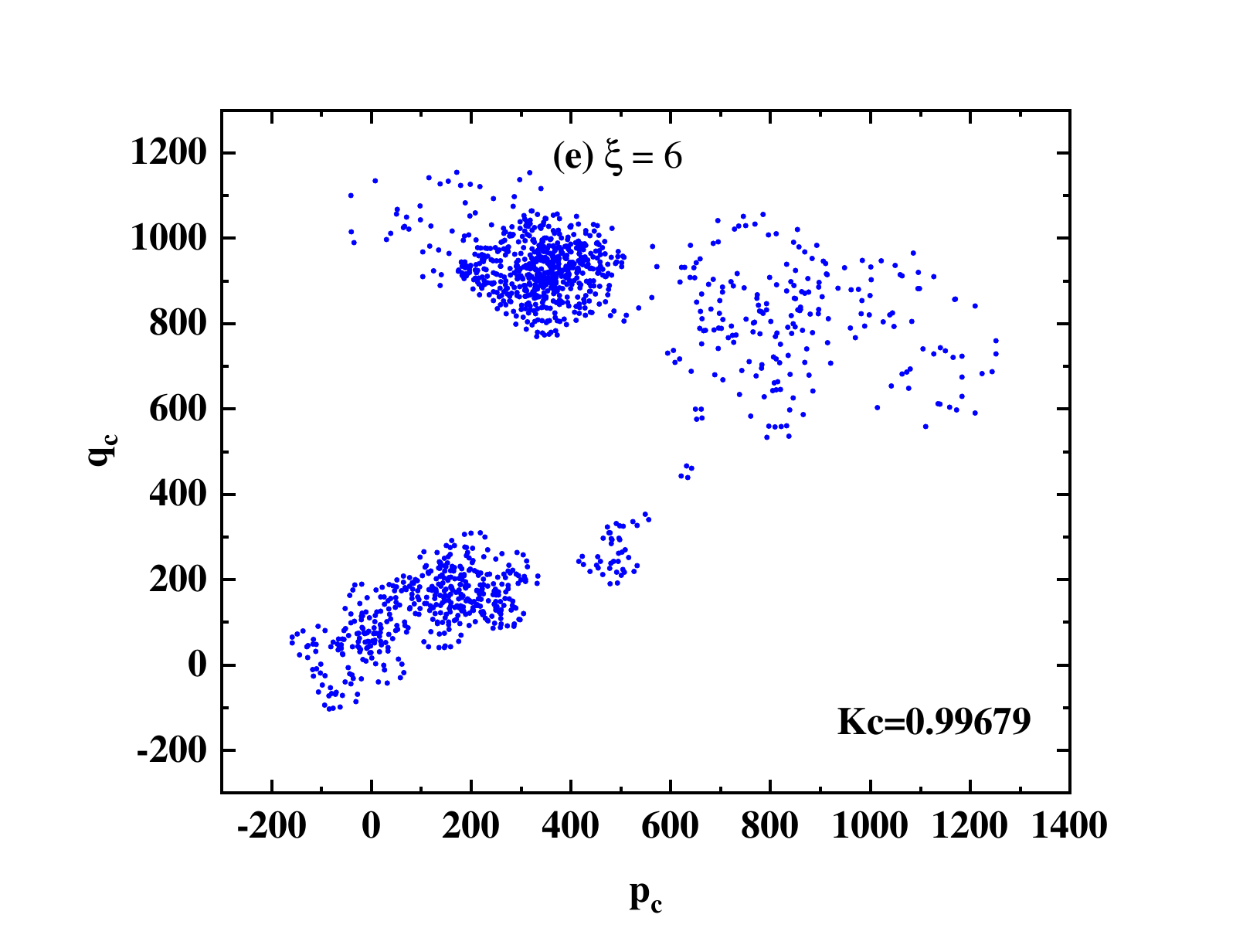}
        \includegraphics[width=17pc]{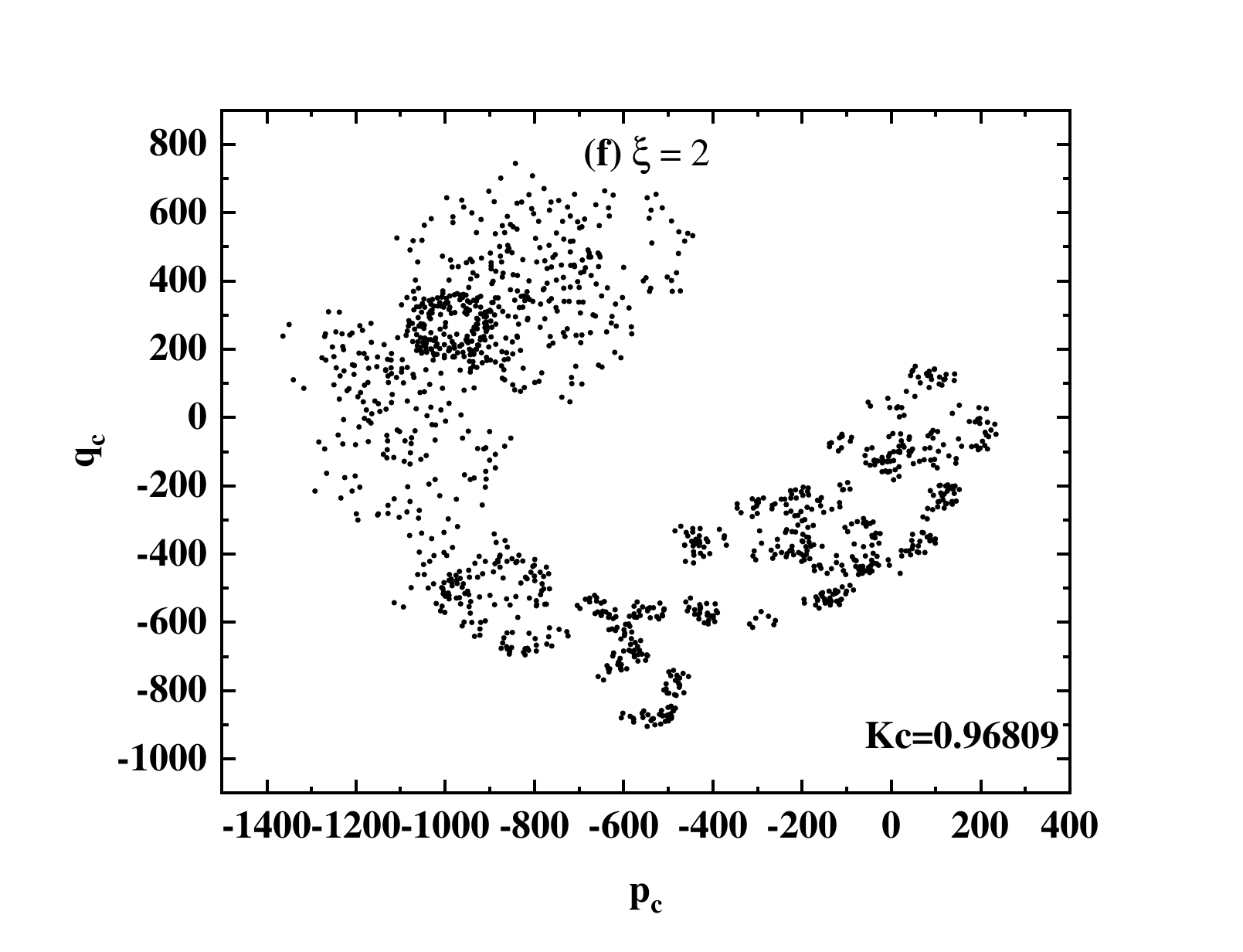}
\caption{Same as Fig. 7 but two values of the parameter $\xi$ in
Fig. 6(d) are adopted.
        }
    }
\end{figure*}

\end{document}